\documentclass[11pt]{amsart}

\usepackage{fullpage}
\usepackage{tikz}
\usepackage{hyperref}
\usepackage[all,cmtip]{xy}
\usepackage{amsopn}
\usepackage{amsfonts}
\usepackage{amsmath}
\usepackage{amsthm}
\usepackage{amscd}
\usepackage{amssymb}
\usepackage{graphicx}

\usetikzlibrary{decorations.markings}

\usepackage{times} 

\makeatletter\@addtoreset {equation}{section}\makeatother

\theoremstyle{plain}

{\begin{trivlist} \item[]{\em Proof }}%
{\hspace*{\fill}$\rule{.3\baselineskip}{.35\baselineskip}$\end{trivlist}}

\begin{document}

\title{Krein Signature in Hamiltonian and $\mathcal{PT}$-symmetric Systems}

\author{A.~Chernyansky}
\address{Department of Mathematics and Statistics, McMaster University, Hamilton, Ontario, Canada, L8S 4K1}
\email{chernya@math.mcmaster.ca}

\author{P.G.~Kevrekidis}
\address{Department of Mathematics and Statistics, University of Massachusetts, Amherst, Massachusetts 01003--4515 USA}
\email{kevrekid@math.umass.edu}

\author{D.E.~Pelinovsky}
\address{Department of Mathematics and Statistics, McMaster University, Hamilton, Ontario, Canada, L8S 4K1}
\email{dmpeli@math.mcmaster.ca}

\date{\today}

\begin{abstract}
We explain the concept of Krein signature in Hamiltonian and $\mathcal{PT}$-symmetric systems
on the case study of the one-dimensional Gross--Pitaevskii equation with a real harmonic potential
and an imaginary linear potential. These potentials correspond to the magnetic trap, and a linear gain/loss
in the mean-field model of cigar-shaped Bose--Einstein condensates.
For the linearized Gross--Pitaevskii equation, we
introduce the real-valued Krein quantity, which is nonzero if the eigenvalue is neutrally stable and simple
and zero if the eigenvalue is unstable. If the neutrally stable eigenvalue is simple, it persists
with respect to perturbations. However, if it is multiple, it may split into unstable eigenvalues under perturbations.
A necessary condition for the onset of instability past the bifurcation point
requires existence of two simple neutrally stable eigenvalues of opposite Krein signatures
before the bifurcation point. This property is useful in the parameter continuations
of neutrally stable eigenvalues of the linearized Gross--Pitaevskii equation.
\end{abstract}

\maketitle


\section{Introduction}
\label{sec-intro}

We consider the prototypical example of the one-dimensional Gross-Pitaevskii (GP) equation
arising in the context of cigar-shaped Bose--Einstein (BEC) condensates~\cite{book1,book2}.
The model takes the form of the following defocusing nonlinear Schr\"{o}dinger (NLS) equation
with a harmonic potential~\cite{emergent,book_new}:
\begin{eqnarray}
i \partial_t u = - \partial_x^2 u + V(x) u + |u|^2 u,
\label{nls}
\end{eqnarray}
where $u$ represents the complex wave function and $V$ characterizes the external potential.
The probability density of finding atoms at a given location and time
is characterized by $|u|^2$.

In the case of magnetic trapping of the BECs~\cite{book1,book2}, the potential $V$ is real-valued and is given by
\begin{equation}
V(x) = \Omega^2 x^2,
\label{pot-Ham}
\end{equation}
where $\Omega$ is the ratio of longitudinal to transverse confinement
strengths of the parabolic trapping.
The NLS equation (\ref{nls}) with the potential (\ref{pot-Ham})
is a Hamiltonian system written in the symplectic form
\begin{equation}
i \frac{\partial u}{\partial t} = \frac{\delta H}{\delta \bar{u}},
\label{symplectic-Ham}
\end{equation}
where $H$ is the following real-valued Hamiltonian function
\begin{equation}
H(u) = \int_{\mathbb{R}} \left[ |\partial_x u|^2 + V(x) |u|^2 + \frac{1}{2} |u|^4 \right] dx.
\label{energy-Ham}
\end{equation}

In the case of effects observed when quantum particles are loaded in an open system,
the external potential $V$ may be complex-valued~\cite{wunner,dast}. The intervals with positive
and negative imaginary part of $V$ correspond to the gain and loss of quantum particles,
respectively. If the gain and loss are modelled by linear functions and the gain matches loss exactly,
the external potential is given by
\begin{equation}
V(x) = \Omega^2 x^2 + 2 i \gamma x,
\label{pot-PT}
\end{equation}
where $\gamma$ is the gain-loss strength. The NLS equation (\ref{nls}) with the potential (\ref{pot-PT})
can still be cast to the symplectic form (\ref{symplectic-Ham}) but the Hamiltonian function $H$ in (\ref{energy-Ham})
is now complex-valued. The complex-conjugate equation to (\ref{symplectic-Ham}) is determined by $\bar{H}$
with $\bar{H} \neq H$. Hence, the NLS equation (\ref{nls}) with the potential (\ref{pot-PT}) is not a Hamiltonian system.

Although $V$ in (\ref{pot-PT}) is not real-valued, it satisfies the following condition
\begin{equation}
\label{potentials}
V(x) = \overline{V(-x)}, \quad x \in \mathbb{R}.
\end{equation}
Let us introduce the parity operator $\mathcal{P}$ and the time reversal operator $\mathcal{T}$ acting on a function $u(x,t)$ as follows:
\begin{equation}
\label{operators}
    \mathcal{P} u(x,t) = u(-x,t), \quad
    \mathcal{T} u(x,t) = \overline{u(x,-t)}.
\end{equation}
Then, we can see that $V$ satisfying (\ref{potentials}) is $\mathcal{PT}$-symmetric under the simultaneous action of
operators (\ref{operators}). We say that the NLS equation (\ref{nls}) with the potential (\ref{pot-PT}) is
$\mathcal{PT}$-symmetric. For any solution $u(x,t)$,
\begin{equation}
\widetilde{u}(x,t) = \mathcal{PT} u(x,t) = \overline{u(-x,-t)}
\end{equation}
is also a solution to the same NLS equation (\ref{nls}) with the potential (\ref{pot-PT}).

Such $\mathcal{PT}$-symmetric models have attracted substantial
attention over the past two decades. They were initially proposed in
the context of a (non-Hermitian) variant of
quantum mechanics \cite{Bender1,Bender2} (see also review in \cite{bender}).
However, their experimental realization in both
low-dimensional (e.g., dimer)~\cite{Ruter} and high-dimensional (e.g.,
lattice)~\cite{ncomms2015} settings have been confirmed in nonlinear optics. This direction
has also inspired an extensive volume of theoretical activity
and even experiments in other areas, including mechanical~\cite{Bender3}
and electrical~\cite{Schindler} systems. Two recent
reviews on the subject can be found in~\cite{Konotop,RevPT}.

The concept of Krein signatures was introduced by MacKay~\cite{mackay} for the
finite-dimensional linear Hamiltonian systems, although the idea dates
back to the works of Weierstrass~\cite{weierstrass}. In the setting
of the NLS equation (\ref{nls}) with the potential (\ref{pot-Ham}),
the linear Hamiltonian system can be formulated as the spectral problem
\begin{equation}
\label{lin-Ham}
J \mathcal{L} v = \lambda v,
\end{equation}
where $\mathcal{L}$ is a self-adjoint unbounded operator in the space of square-integrable
functions $L^2(\mathbb{R})$ with a dense domain in $L^2(\mathbb{R})$
and $J$ is a skew-adjoint bounded operator in $L^2(\mathbb{R})$.
The operators $\mathcal{L}$ and $J$ are assumed to satisfy
$J^2 = -I$ and $J\mathcal{L} + \mathcal{\bar{L}} \bar{J} = 0$,
thanks to the Hamiltonian symmetry.

If $\lambda_0 \in \mathbb{C}$ is an eigenvalue of the spectral problem (\ref{lin-Ham}),
then it is neutrally stable if ${\rm Re}(\lambda_0) = 0$ and unstable
if ${\rm Re}(\lambda_0) > 0$. Thanks to the Hamiltonian symmetry of $\mathcal{L}$ and $J$,
the eigenvalues appear in symmetric pairs relative to the axis ${\rm Re}(\lambda) = 0$.
Indeed, if $v$ is an eigenvector of the spectral problem \eqref{lin-Ham}
for the eigenvalue $\lambda$, then $w = - J \bar{v}$ is an eigenvector of the same
spectral problem (\ref{lin-Ham}) with the eigenvalue $-\bar{\lambda}$. Indeed,
substituting $v = \bar{J} \bar{w}$ into (\ref{lin-Ham}) yields
\begin{equation*}
J \mathcal{L} \bar{J} \bar{w}  = \lambda \bar{J} \bar{w} \quad \Leftrightarrow \quad
\mathcal{\bar{L}} \bar{w} = \lambda \bar{J} \bar{w} \quad \Leftrightarrow \quad
\bar{J} \mathcal{\bar{L}} \bar{w}  = -\lambda \bar{w} \quad \Leftrightarrow \quad
J \mathcal{L} w  = -\bar{\lambda} w.
\end{equation*}

For a nonzero eigenvalue $\lambda_0 \in \mathbb{C}$ of the spectral problem (\ref{lin-Ham})
with the eigenvector $v_0$ in the domain of $\mathcal{L}$,
we define the Krein quantity $K(\lambda_0)$ by
\begin{equation}
K(\lambda_0) := \langle \mathcal{L} v_0, v_0 \rangle,
\label{krein-intro}
\end{equation}
where $\langle \cdot, \cdot \rangle$ is the standard inner product in $L^2(\mathbb{R})$.
Krein quantity in (\ref{krein-intro}) satisfies the following properties:

\vspace{2pt}
\centerline{\fbox{\parbox[cs]{0.6\textwidth}{
\begin{enumerate}
\item $K(\lambda_0)$ is real if $\lambda_0 \in i \mathbb{R}$.
\item $K(\lambda_0)$ is nonzero if $\lambda_0 \in i \mathbb{R} \backslash \{0\}$ is simple.
\item $K(\lambda_0)$ is zero if $\lambda_0 \in \mathbb{C} \backslash \{ i \mathbb{R} \}$.
\end{enumerate}
}}}
\vspace{2pt}
The Krein signature is defined as the sign of the Krein quantity $K(\lambda_0)$ for
a simple neutrally stable eigenvalue $\lambda_0 \in i \mathbb{R} \backslash \{0\}$.
If parameters of the NLS equation (\ref{nls}) change, parameters
of the spectral problem (\ref{lin-Ham}) change, however, the simple eigenvalue
$\lambda_0 \in i \mathbb{R}$ remains on the axis ${\rm Re}(\lambda) = 0$
unless it coalesces with another eigenvalue or a part of the continuous spectrum,
thanks to the preservation of its multiplicity and the Hamiltonian symmetry of eigenvalues.
In this case, the eigenvalue $\lambda_0$ and its Krein quantity $K(\lambda_0)$
are at least continuous functions of the parameters of the NLS equation (\ref{nls}).

It is quite typical in the parameter continuations of the spectral problem (\ref{lin-Ham})
to see that the simple eigenvalue $\lambda_0 \in i \mathbb{R}$ coalesces at a bifurcation
point with another simple eigenvalue $\lambda_0' \in i \mathbb{R}$ and that both eigenvalues
split into the complex plane as unstable eigenvalues past the bifurcation point.
The Krein signature is a helpful tool towards  predicting this instability bifurcation in the sense
of the following necessary condition.

\vspace{2pt}
\centerline{\fbox{\parbox[cs]{0.8\textwidth}{
{\bf Necessary condition for instability bifurcation.} {\em Under some non-degeneracy constraints, the double eigenvalue $\lambda_0 = \lambda_0'
\in i \mathbb{R}$ of the spectral problem (\ref{lin-Ham}) with a bifurcation parameter $\varepsilon \in \mathbb{R}$
splits into a pair of complex eigenvalues symmetric relative to ${\rm Re}(\lambda) = 0$ for $\varepsilon > 0$
only if there exist two simple eigenvalues $\lambda_0, \lambda_0' \in i \mathbb{R}$ with the opposite
Krein signature for $\varepsilon < 0$.}
}}}
\vspace{2pt}

In other words, if two neutrally stable eigenvalues of the same Krein signature move towards each other
in the parameter continuation of the spectral problem (\ref{lin-Ham}), then their coalescence
will not result in the onset of instability, whereas if the two neutrally stable eigenvalues have
the opposite Krein signature, their coalescence is likely to result in the onset of
instability, subject to technical non-degeneracy constraints.

The concept of Krein signature in the infinite-dimensional setting, e.g. for the NLS equation,
was introduced independently in works \cite{KKS,Pel05}. It was justified in a number of mathematical
publications \cite{CP10,KP10} and it remains a practical tool to trace instability bifurcations
in physically relevant Hamiltonian systems \cite{PelYang,skryabin} (see review in \cite{Kollar}).
In particular, the following completeness result is available for the Hamiltonian systems.

\vspace{2pt}
\centerline{\fbox{\parbox[cs]{0.8\textwidth}{
      {\bf Hamiltonian--Krein Theorem.} {\em If $\mathcal{L}$ has no kernel, has
        finitely many negative eigenvalues
$n(\mathcal{L}) < \infty$, and the rest of its spectrum is strictly positive,
then eigenvalues of the spectral problem (\ref{lin-Ham}) satisfy the completeness relation
$$
n(\mathcal{L}) = N_{\rm real} + N_{\rm comp} + N_{\rm imag}^-,
$$
where $N_{\rm real}$ is the number of real positive eigenvalues $\lambda$,
$N_{\rm comp}$ is the number of complex eigenvalues $\lambda$ with ${\rm Re}(\lambda) > 0$,
and $N_{\rm imag}^-$ is the number of purely imaginary eigenvalues $\lambda$ with negative
Krein signature, respectively. All numbers are accounted in their algebraic multiplicity.}
}}}
\vspace{5pt}

In the context of the NLS equation (\ref{nls}) with the potential (\ref{pot-Ham}), the phase invariance introduces
a symmetry and a kernel of the operator $\mathcal{L}$. In this case, the negative index $n(\mathcal{L})$
has to be recomputed in a subspace of $L^2(\mathbb{R})$ which is $J$-orthogonal to the kernel of $\mathcal{L}$.
See monographs \cite{kap-book,Pel-book} for further mathematical details.

It was only very recently that the concept of Krein signature was extended
to the non-Hamiltonian $\mathcal{PT}$-symmetric systems. The linear Schr\"{o}dinger equation
with a complex-valued $\mathcal{PT}$-symmetric potential was considered in \cite{yang}, where
the indefinite $\mathcal{PT}$-inner product with the induced $\mathcal{PT}$-Krein signature was introduced
in the exact correspondence with the Krein signature for the Hamiltonian spectral problem (\ref{lin-Ham}).
Coupled non-Hamiltonian $\mathcal{PT}$-symmetric systems
with constant coefficients were considered in \cite{AB,ABdiscrete} (see also \cite{SS}),
where the linearized problem was block-diagonalized to the form for which
the Krein signature of eigenvalues can be introduced.
A Hamiltonian version of the $\mathcal{PT}$-symmetric system of coupled oscillators was considered
in \cite{CP1,CP2}, where the Krein signature of eigenvalues was introduced by using the corresponding Hamiltonian.
Finally, Krein signature of eigenvalues was defined in \cite{ChernPel} for the
spectral problem related to the linearization of the NLS equation with complex-valued potentials.

Compared to the Hamiltonian case in \cite{KKS,Pel05}
and to the linear $\mathcal{PT}$-symmetric Schr\"{o}dinger equation in \cite{yang},
it was shown in \cite{ChernPel} that the Krein signature
of eigenvalues in the linearization of the $\mathcal{PT}$-symmetric NLS equation
cannot be computed just from the eigenvectors in the spectral
problem. This is because the adjoint eigenvectors need to be computed
separately and the sign of the adjoint eigenvector needs to be chosen by a continuity argument.
This limits practical applications of the Krein signature in nonlinear $\mathcal{PT}$-symmetric
systems. Nevertheless, all the main definitions and properties of the Krein quantity
listed above for the Hamiltonian NLS equation are extended to the case of the $\mathcal{PT}$-symmetric NLS equation.
Moreover, the necessary condition for the instability bifurcation is extended to the $\mathcal{PT}$-symmetric
NLS equation but not the Hamiltonian--Krein Theorem.

The purpose of this chapter is to explain definitions and properties of the Krein signature
on the prototypical example of the NLS equation (\ref{nls}) with either the potential
(\ref{pot-Ham}) or the potential (\ref{pot-PT}).

We also address the Krein signature for the linear $\mathcal{PT}$-symmetric Schr\"{o}dinger equation
as the one introduced in \cite{yang}, where we discuss differences from the Krein signature
in the linearized $\mathcal{PT}$-symmetric NLS equation. We will show that the linear Schr\"{o}dinger
equation with a real even potential (\ref{pot-Ham}) admits two equivalent Hamiltonian formulations
and hence two equivalent definitions of the Krein signatures. The standard Hamiltonian formulation
leads to eigenvalues of only positive Krein signature, whereas the non-standard Hamiltonian formulation
leads to infinitely many eigenvalues of opposite Krein signature. It is the latter Hamiltonian formulation
that can be extended to the case of the $\mathcal{PT}$-symmetric potential (\ref{pot-PT}).

This chapter is organized as follows. Section \ref{sec-Ham} addresses nonlinear stationary states
bifurcating from simple eigenvalues of the quantum harmonic oscillator and describes Krein signature
in the linearized NLS equation with the potential (\ref{pot-Ham}). Section
\ref{sec-PT} describes Krein signature for the $\mathcal{PT}$-symmetric NLS equation with the potential (\ref{pot-PT}),
where we highlight the differences between the Hamiltonian and the $\mathcal{PT}$-symmetric cases.
Section \ref{sec-lin-PT} contains discussion of the linear $\mathcal{PT}$-symmetric Schr\"{o}dinger equation.
Section \ref{sec-conclusion} summarizes the results and lists further directions.

\section{Krein signature for the NLS equation}
\label{sec-Ham}

In the context of the NLS equation (\ref{nls}) with the potential (\ref{pot-Ham}),
we consider the nonlinear stationary states of the form $u(x,t) = e^{-i \mu t} \phi(x)$,
where $\mu \in \mathbb{R}$ is referred to as the chemical potential~\cite{dast} and
the real-valued function $\phi$ satisfies the differential equation
\begin{eqnarray}
\mu \phi(x) = - \phi''(x) + x^2 \phi(x) + \phi(x)^3,
\label{nls-stat}
\end{eqnarray}
where we have set $\Omega = 1$ without loss of generality. In the linear (small-amplitude)
limit, we obtain the quantum harmonic oscillator with the eigenvalues $\mu_n = 1 + 2n$,
$n \in \mathbb{N}_0 := \{0,1,2,...\}$ and the $L^2$-normalized eigenfunctions
\begin{equation}
\label{eigenvector}
\varphi_n(x) = \frac{1}{\sqrt{2^n n! \sqrt{\pi}}} H_n(x) e^{-x^2/2},
\end{equation}
where $H_n$ is the Hermite polynomial of degree $n$, e.g.,
$H_0(x) = 1$, $H_1(x) = 2x$, $H_2(x) = 4x^2 - 2$, etc.

Each eigenfunction $\varphi_n$ for a simple eigenvalue $\mu_n$ generates a
branch of solutions bifurcating in the stationary problem (\ref{nls-stat}).
This follows from the general Crandall--Rabinowitz bifurcation theory \cite{CR}
and is generally used in physics community, see, e.g.,~\cite{mprizola,Alfimov}.
Each branch can be approximated by the following expansion in terms
of the small parameter $\epsilon$:
\begin{equation}
\label{expansion}
\left\{ \begin{array}{l}
\mu = \mu_n + \epsilon^2 \mu_n^{(2)} + \dots, \\
\phi = \epsilon \varphi_n + \epsilon^{3} \varphi_n^{(3)} + \dots,
\end{array} \right.
\end{equation}
where $(\mu_n,\varphi_n)$ is the $n$-th eigenvalue--eigenfunction pair,
$(\mu_n^{(2)},\varphi_n^{(3)})$ are the next-order correction terms to be found,
and the dots denote the higher-order corrections terms.
The $n$-th branch of the nonlinear stationary states is smooth with respect to
the small parameter $\epsilon$, which parameterizes both $\mu$ and $\phi$, whereas it
has a square-root singularity when it is written in terms of the parameter $\mu - \mu_n$.

The formal solvability condition for the correction terms
$(\mu_n^{(2)},\varphi_n^{(3)})$ yields
\begin{eqnarray}
\mu_n^{(2)} =  \int_{\mathbb{R}} \varphi_n(x)^4 dx  > 0,
\label{eqn2}
\end{eqnarray}
which implies that the branch of nonlinear stationary states extends towards $\mu > \mu_n$.
The limit $\mu \to \infty$ can be rescaled as the semi-classical limit of the stationary
NLS equation. Each $n$-th branch of the nonlinear stationary states is uniquely extended
to the limit $\mu \to \infty$, where it is matched with the asymptotic approximation
involving bound states of $n$ dark solitons on the background of $V$ in (\ref{pot-Ham})
\cite{coles,Pelin}.

When considering the stability of the nonlinear stationary state of the form $u(x,t) = e^{-i \mu t} \phi(x)$,
we linearize the NLS equation (\ref{nls}) with the expansion
\begin{eqnarray}
u(x,t)= e^{-i \mu t} \left[ \phi(x) + \delta
\left( a(x) e^{-\lambda t} + \bar{b}(x) e^{-\bar{\lambda} t} \right)  + \dots \right],
\label{eqn1}
\end{eqnarray}
where $\delta$ is a formal small parameter. To the leading
order in $\delta$, the eigenvalue--eigenvector pair $(\lambda,v)$ with $v = (a,b)^T$
is found from the spectral problem
\begin{eqnarray}
\mathcal{L} v = -i \lambda \sigma_3 v,
\label{spectrum}
\end{eqnarray}
where $\sigma_3 = {\rm diag}(1,-1)$ and the linear operator $\mathcal{L}$ is written in the differential form:
\begin{eqnarray}
\mathcal{L} = \left[ \begin{array}{cc} -\partial_x^2 + x^2 - \mu + 2 \phi(x)^2 & \phi(x)^2 \\
    \phi(x)^2 & -\partial_x^2 + x^2 - \mu + 2 \phi(x)^2 \end{array} \right].
\label{hh}
\end{eqnarray}
The operator $\mathcal{L}$ is extended to a self-adjoint operator in $L^2(\mathbb{R})$
with the domain $H^2(\mathbb{R}) \cap L^{2,2}(\mathbb{R})$ (see ~\cite{heffler}, Ch. 4, p.37),
where $H^2(\mathbb{R})$ is the Sobolev space of square integrable functions and their
second derivatives and $L^{2,2}(\mathbb{R})$ is the space of square integrable
functions multiplied by $(1+x^2)$. The spectrum of $\mathcal{L}$ is purely discrete
(see~\cite{reed4}, Ch. XIII, Theorem 16 on p.120).

The spectral problem (\ref{spectrum}) takes the abstract form
(\ref{lin-Ham}) with the self-adjoint operator $\mathcal{L}$ given by (\ref{hh}) and
the skew-symmetric operator $J = i\sigma_3$. The Hamiltonian symmetry $J^2 = - I$
and $J\mathcal{L} + \mathcal{\bar{L}} \bar{J} = 0$ (or, equivalently, $\sigma_3 \mathcal{L} = \mathcal{\bar{L}} \sigma_3$)
is satisfied. The eigenvalues are symmetric relative to the imaginary axis.
To be precise, if $\lambda_0$ is an eigenvalue with the eigenvector $v_0 = (a,b)^T$,
then $-\overline{\lambda}_0$ is another eigenvalue with the eigenvector
$\sigma_3 \bar{v}_0 = (\overline{a},-\overline{b})^T$
by the Hamiltonian symmetry $\sigma_3 \mathcal{L} = \mathcal{\bar{L}} \sigma_3$.

In addition to the Hamiltonian symmetry, the operator $\mathcal{L}$ in (\ref{hh})
satisfies $\sigma_1 \mathcal{L} = \mathcal{\bar{L}} \sigma_1$, which implies that
the eigenvalues are symmetric relative to the real axis. Indeed,
if $\lambda_0$ is an eigenvalue with the eigenvector $v_0 = (a,b)^T$, then
$\bar{\lambda}_0$ is another eigenvalue with the eigenvector $\sigma_1 \bar{v}_0 = (\bar{b},\bar{a})$.
Hence, the unstable eigenvalues with ${\rm Re}(\lambda_0) > 0$
occur either as pairs on the real axis or as quadruplets in the complex plane,
whereas the neutrally stable eigenvalues with ${\rm Re}(\lambda_0) = 0$
occur as pairs on the imaginary axis.

For each nonzero eigenvalue $\lambda_0 \in \mathbb{C}$ of the spectral problem
(\ref{spectrum}) with the eigenvector $v_0 = (a,b)^T \in H^2(\mathbb{R}) \cap L^{2,2}(\mathbb{R})$,
the Krein quantity $K(\lambda_0)$ introduced in (\ref{krein-intro}) can be written
explicitly as follows:
\begin{eqnarray}
K(\lambda_0) = \langle \mathcal{L} v_0, v_0 \rangle = -i \lambda_0 \langle \sigma_3 v_0, v_0 \rangle
= -i \lambda_0 \int_{\mathbb{R}} (|a(x)|^2 - |b(x)|^2) dx.
\label{eqn2b}
\end{eqnarray}
If $K(\lambda_0)$ is nonzero and real, the sign of $K(\lambda_0)$ is referred to as
the Krein signature. In what follows, we only consider eigenvalues with $\lambda_0 \in i \mathbb{R}_+$,
for which $-i \lambda_0 > 0$.

Let us verify the three main properties of the Krein quantity $K(\lambda_0)$.
\begin{enumerate}
\item If $\lambda_0 \in i\mathbb{R}$, then $(-i\lambda_0) \in \mathbb{R}$.
The integral in ~\eqref{eqn2b} is also real. Hence, $K(\lambda_0)$ is real.

\item Let us write the eigenvalue problem~\eqref{hh} for the generalized eigenvector $v_g$:
\begin{equation}
(\mathcal{L} + i\lambda_0 \sigma_3)v_g = \sigma_3 v_0.
\label{hhg}
\end{equation}
If $\lambda_0 \in i\mathbb{R}\backslash \{0\}$, then $v_0$ is in the kernel of the adjoint operator
$(\mathcal{L} + i\lambda_0\sigma_3)^*$, and
Fredholm solvability condition of the above equation is $\langle \sigma_3 v_0,v_0\rangle = 0$.
If $K(\lambda_0) = 0$, then there exists a solution to the nonhomogeneous equation (\ref{hhg}),
so that $\lambda_0$ is not simple. Hence, $K(\lambda_0) \neq 0$.

\item Using self-adjoint property of $\mathcal{L}$, one can write
\begin{equation*}
\langle \mathcal{L} v_0, v_0\rangle =
\langle v_0, \mathcal{L} v_0\rangle,
\end{equation*}
which can be expanded as
\begin{equation*}
-i\lambda_0 \langle \sigma_3 v_0,v_0\rangle = i\bar\lambda_0 \langle v_0,\sigma_3 v_0\rangle,
\end{equation*}
where the equality holds either for $\lambda_0\in i\mathbb{R}$ or $\langle \sigma_3 v_0,v_0\rangle = 0$.
Hence $K(\lambda_0)=0$ for $\lambda_0 \not\in i\mathbb{R}$.
\end{enumerate}

Let us now illustrate how the Krein signatures can be used to predict instability bifurcations
from multiple neutrally stable eigenvalues of the spectral problem (\ref{spectrum}).
We restrict consideration to the small-amplitude limit.
If $\epsilon = 0$ and $\mu = \mu_n$, the linear operator (\ref{hh}) becomes diagonal:
\begin{eqnarray}
\mathcal{L}_0 = \left[ \begin{array}{cc} -\partial_x^2 + x^2 - \mu_n & 0 \\
    0 & -\partial_x^2 + x^2 - \mu_n \end{array} \right]
\end{eqnarray}
and the eigenvalues are located at $\sigma(\mathcal{L}_0) = \{ 2(m-n), \;\; m \in \mathbb{N}_0 \}$,
where $n \in \mathbb{N}_0$ is fixed. Because of the skew-symmetric operator $J = i\sigma_3$ in the right-hand side of
the spectral problem (\ref{spectrum}), these eigenvalues are mapped to the imaginary
axis in the pairs $\lambda \in \pm i \{ 2(m-n), \; m \in \mathbb{N}_0\}$.

If $n = 0$, the ground state branch (\ref{expansion}) leads to a double zero eigenvalue
and a set of simple eigenvalues in pairs $\lambda \in \pm i \{ 2m, \; m \in \mathbb{N}_0 \backslash \{0\}\}$.
The double zero eigenvalue is preserved in $\epsilon$ due to gauge symmetry, whereas
the simple neutrally stable eigenvalues are preserved on the imaginary axis due to Hamiltonian symmetry
(at least for small $\epsilon$). Moreover, each eigenvalue has a positive Krein signature,
therefore, by the necessary condition for instability bifurcations, no
complex eigenvalue quartets can arise in parameter continuations of solutions
to the spectral problem (\ref{spectrum}) in $\epsilon$. These spectral stability properties are natural
for the ground state solution.

If $n = 1$, the first excited state branch (\ref{expansion}) associated
with a single dark soliton \cite{coles,Pelin} leads to a double zero eigenvalue,
a pair of double eigenvalues $\lambda = \pm 2i$, and a set of simple eigenvalues in pairs
$\lambda \in \pm i \{ 2 (m-1), \; m \in \mathbb{N}_0 \backslash \{0, \pm 1 \}\}$.
The double zero eigenvalue is again preserved in $\epsilon$ due to gauge symmetry but the pair
of nonzero double eigenvalues $\lambda = \pm 2i$ may split if $\epsilon \neq 0$.
Note that two linearly independent eigenvectors exist for $\lambda_0 = 2i$:
\begin{equation}
\label{eigenvectors-first}
v_1 = \left[ \begin{array}{c} \varphi_2 \\ 0 \end{array} \right], \quad
v_2 = \left[ \begin{array}{c} 0 \\ \varphi_0 \end{array} \right].
\end{equation}
The two eigenvectors induce opposite Krein signatures for the coalescent double eigenvalue
since $K(\lambda_0) > 0$ for $v_1$ and $K(\lambda_0) < 0$ for $v_2$. Therefore,
by the necessary condition on the splitting of the double eigenvalues, we may anticipate
unstable eigenvalues for small $\epsilon$.

Similarly, if $n = 2$, the second excited state branch (\ref{expansion}) associated
with two dark solitons \cite{coles,Pelin} leads to a double zero eigenvalue,
two pairs of double eigenvalues $\lambda = \pm 2i$ and $\lambda = \pm 4 i$, and a set of simple eigenvalues in pairs
$\lambda \in \pm i \{ 2 (m-2), \; m \in \mathbb{N}_0 \backslash \{0, \pm 1, \pm 2 \}\}$.
The double zero eigenvalue is again preserved in $\epsilon$ due to gauge symmetry but the pairs
of nonzero double eigenvalues $\lambda = \pm 2i$ and $\lambda = \pm 4i$ may split if $\epsilon \neq 0$.
Note that two linearly independent eigenvectors exist as follows:
\begin{equation}
\label{eigenvectors-second-1}
\lambda_0 = 2i : \quad
v_1 = \left[ \begin{array}{c} \varphi_3 \\ 0 \end{array} \right], \quad
v_2 = \left[ \begin{array}{c} 0 \\ \varphi_1 \end{array} \right]
\end{equation}
and
\begin{equation}
\label{eigenvectors-second-2}
\lambda_0 = 4i : \quad
v_1 = \left[ \begin{array}{c} \varphi_4 \\ 0 \end{array} \right], \quad
v_2 = \left[ \begin{array}{c} 0 \\ \varphi_0 \end{array} \right].
\end{equation}
Again, the two eigenvectors induce opposite Krein signatures for each coalescent double eigenvalue, hence
by the necessary condition on the splitting of the double eigenvalues,
we may anticipate unstable eigenvalues for small $\epsilon$.

In order to compute definite predictions whether or not the double
eigenvalues produce instability bifurcations for the first and second excited states,
we shall proceed using perturbation
theory arguments. We substitute expansion (\ref{expansion}) into the spectral problem (\ref{spectrum})
and expand it into powers of $\epsilon^2$ as follows:
\begin{equation}
\label{perturbed-spectrum}
(\mathcal{L}_0 + \epsilon^2 \mathcal{L}_1 + \dots) v = - i \lambda \sigma_3 v,
\end{equation}
where
\begin{eqnarray}
\label{Eq:uv2}
\mathcal{L}_1 = \left[ \begin{array}{cc}
2  \varphi_n(x)^2 - \mu_n^{(2)} &  \varphi_n(x)^2 \\
\varphi_n(x)^2  & 2 \varphi_n(x)^2 - \mu_n^{(2)} \end{array} \right].
\end{eqnarray}

Let $-i\lambda = \omega_0 + \epsilon^2 \omega_1 + \dots$, where $\omega_0$
is a coalescent double eigenvalue and $\omega_1$ is a correction term.
Representing $v = c_1 v_1 + c_2 v_2 + \dots$ and projecting
the perturbed spectral problem (\ref{perturbed-spectrum}) to the eigenvectors
$v_1$ and $v_2$ yield the matrix eigenvalue problem
\begin{equation}
M \left[ \begin{array}{c} c_1 \\ c_2 \end{array} \right] = \omega_1 \sigma_3 \left[ \begin{array}{c} c_1 \\ c_2 \end{array} \right],
\label{eqnM}
\end{equation}
where $M_{ij} = \langle \mathcal{L}_1 v_i, v_j \rangle$, $1 \leq i,j \leq 2$, and
the $L^2$ normalization of eigenvectors has been taken into account.

Let us consider the first excited state $n = 1$ bifurcating from $\mu_1 = 3$.
For $\epsilon = 0$, the eigenvalue at $\omega_0=2$ is double with two eigenvectors (\ref{eigenvectors-first}).
However, there exists a linear combination of $v_1$ and $v_2$ which produces the so-called
dipolar oscillation (also known as the Kohn mode, see explicit solutions in \cite{pelirecent})
and thus the eigenvalue at $\omega_0 = 2$ related to this linear combination
is independent of the variations of the chemical potential in $\epsilon$.
The shift of the eigenvalue for another linear combination of $v_1$ and $v_2$
has been the subject of intense scrutiny as it is associated with the oscillation frequency
of the dark soliton in the parabolic trap~\cite{busch,franz}.

\begin{figure}[htb]
\begin{center}
\includegraphics[width=8cm]{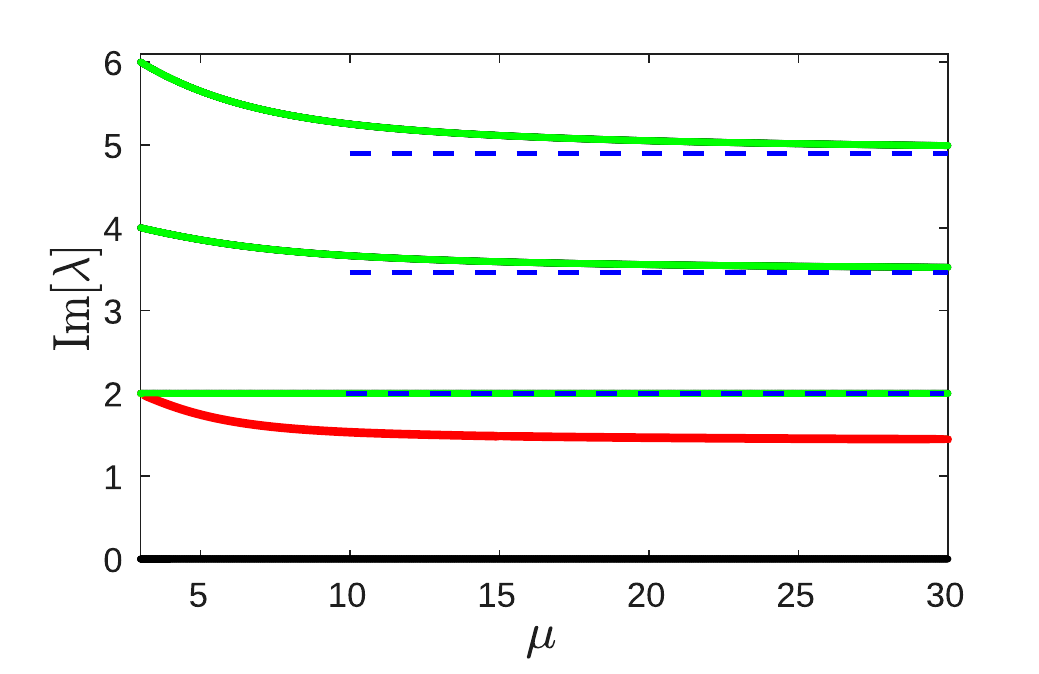}
\includegraphics[width=8cm]{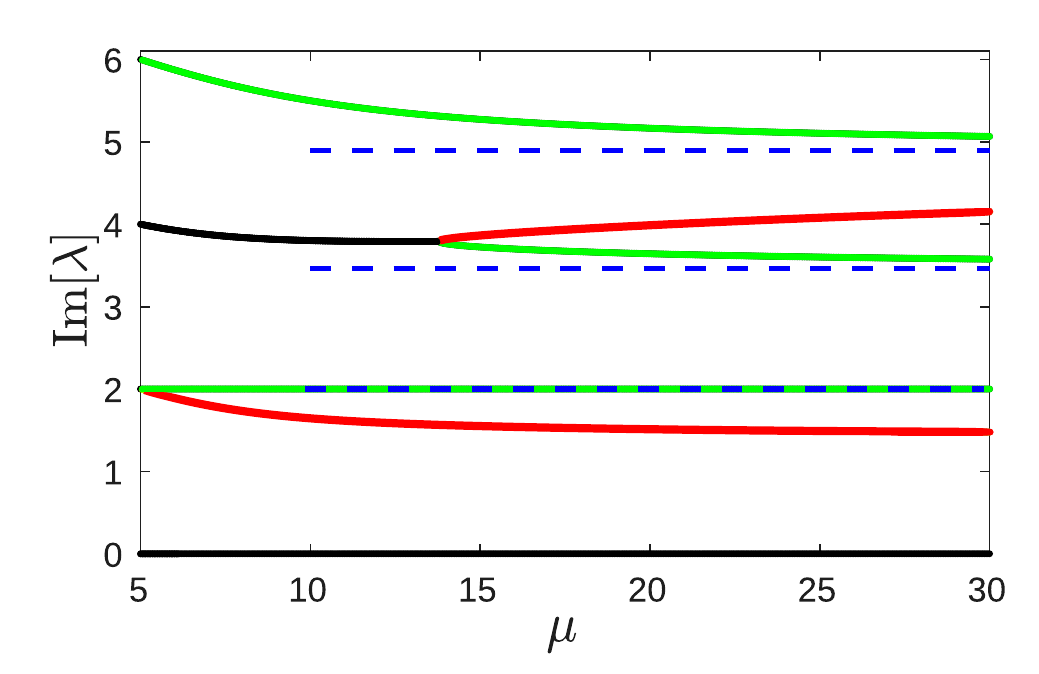}
\includegraphics[width=8cm]{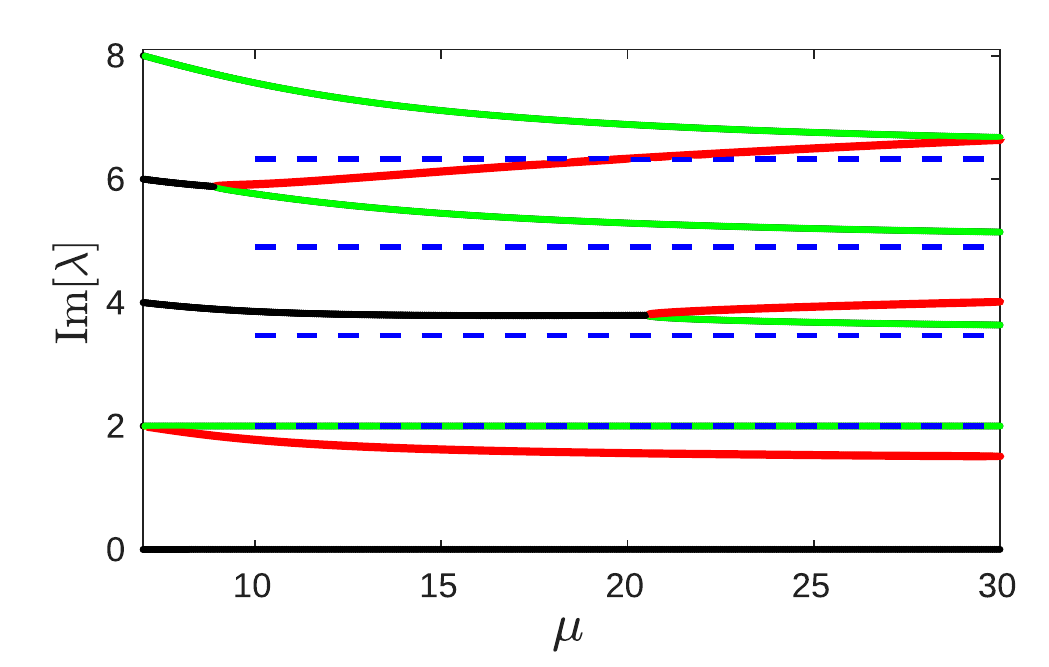}
\caption{The top left panel corresponds to the case of
the first excited state, the top right one corresponds to
the second excited state, while the bottom panel corresponds
to the third excited state. Eigenvalues of negative (positive)
Krein signature are shown in red (green), complex eigenvalues
are shown in black. For the first excited state, only the lowest
  nonzero eigenfrequency has a negative Krein signature
  (but its linear degeneracy with a symmetry mode yields
  no instability). For the second excited state, there are two degenerate
  modes at $2$ and $4$. Only the latter yields the quartet of
  complex eigenvalues. For the third excited states, there are three
  degenerate modes at $2$, $4$, and $6$, the last two yield quartets of complex
  eigenvalues.   }
\label{fig1}
\end{center}
\end{figure}

By using (\ref{eqn2}) for $n = 1$, we find $\mu_1^{(2)} = 3 /(4\sqrt{2\pi})$.
The matrix $M$ in the matrix eigenvalue problem~\eqref{eqnM} is computed explicitly as
\begin{equation}
M = \left[ \begin{array}{cc}
\frac{1}{8\sqrt{2\pi}} & \frac{1}{8\sqrt{\pi}} \\
\frac{1}{8\sqrt{\pi}} & \frac{1}{4\sqrt{2\pi}}
\end{array}\right].
\end{equation}
Computations of eigenvalues of the matrix eigenvalue problem ~\eqref{eqnM} yield $0$ and $-1/(8\sqrt{2\pi})$.
The zero eigenvalue corresponds to the dipolar oscillations. The nonzero eigenvalue near $\omega_0 = 2$ 
is given by the following expansion:
\begin{eqnarray}
\omega= 2 - \frac{1}{6} \left(\mu- 3 \right) + \dots
\label{pred1}
\end{eqnarray}
Numerical results on the top left panel of Figure \ref{fig1} confirm this prediction. The smallest nonzero eigenvalue
remains below $\omega_0 = 2$ and approaches $\omega \rightarrow \sqrt{2}$ as $\mu \to \infty$,
in agreement with the previous results \cite{busch,franz}.

It is relevant to indicate that the asymptotic limit of the eigenfrequencies of the ground state solution
with $n = 0$ can be computed in the limit of large $\mu$ \cite{sandro2} (see also~\cite{pelirecent} for a recent
account of the relevant analysis). These modes include the so-called dipolar oscillation, quadrupolar oscillation, etc.
(associated, respectively, to $m=1$, $m=2$, etc.)
and the corresponding eigenfrequencies are given by
the analytical expression in the limit $\mu \to \infty$:
\begin{eqnarray}
\omega_m = \sqrt{2m (m+1)}, \quad m \in \mathbb{N}.
\label{tflimit}
\end{eqnarray}
We can see from the top left panel of Fig.~\ref{fig1} that these frequencies of the
ground state solution are present in the linearization of the first excited state
in addition to the eigenfrequency $\omega_* = \sqrt{2}$, which corresponds to
the oscillation of the dark soliton inside the trap.

While the example of the first excited state is instructive, it does not
show any instability bifurcations due to coalescence of eigenvalues
of the opposite Krein signatures. This is because although the eigenfrequency at $\omega_0=2$ is double,
the dipolar oscillations do not allow the manifestation of
an instability as a result of resonance. However, the onset of instability
can still be found for the other excited states, e.g. for the second excited state
corresponding to $n = 2$ bifurcating out of $\mu_2 = 5$.

By using (\ref{eqn2}) for $n = 2$, we find $\mu_2^{(2)} = 41 / (64\sqrt{2\pi})$.
At $\epsilon = 0$, the eigenvalue at $\omega_0 = 2$ is double with
the two eigenvectors (\ref{eigenvectors-second-1}).
The dipolar oscillation mode is present again and corresponds to the eigenvalue at $\omega_0 = 2$
independently of the variations of the chemical potential in $\epsilon$.
The other eigenvalue at $\omega_0 = 2$ is shifted for small $\epsilon$.
The matrix $M$ in the matrix eigenvalue problem~\eqref{eqnM} is computed explicitly as
\begin{equation}
M = \left[ \begin{array}{cc}
\frac{5}{32\sqrt{2\pi}} & \frac{15}{64\sqrt{3\pi}} \\
\frac{15}{64\sqrt{3\pi}} & \frac{15}{64\sqrt{2\pi}}
\end{array}\right].
\end{equation}
Computations of eigenvalues of the matrix eigenvalue problem ~\eqref{eqnM} yield $0$ and $-5/(64\sqrt{2\pi})$.
The nonzero eigenvalue near $\omega_0 = 2$ is given by the following expansion:
\begin{eqnarray}
\omega = 2 - \frac{5}{41} \left(\mu - 5\right) + \dots
\label{pred4}
\end{eqnarray}

While the degeneracy at $\omega_0=2$ does not lead to the onset of instability,
let us consider the double eigenvalue at $\omega_0 = 4$ with
the two eigenvectors (\ref{eigenvectors-second-2}). The matrix $M$
in the matrix eigenvalue problem~\eqref{eqnM} is computed explicitly as
\begin{equation}
M = \left[ \begin{array}{cc}
\frac{1}{512\sqrt{2\pi}} & \frac{9}{128\sqrt{3\pi}} \\[5pt]
\frac{9}{128\sqrt{3\pi}} & \frac{7}{64\sqrt{2\pi}}
\end{array}\right].
\end{equation}
The complex eigenvalues of the matrix eigenvalue problem (\ref{eqnM}) are
given by $(-55 \pm 3\sqrt{23}i)/(2048\sqrt{2\pi})$. The complex eigenvalues 
near $\omega_0 = 4$ are given by the following expansion:
\begin{eqnarray}
\omega= 4 + \frac{-55 \pm 3 \sqrt{23} i }{656} (\mu - 5) + \dots
\label{pred5}
\end{eqnarray}
The eigenvalues remain complex for values of $\mu \gtrsim 5$ but coalesce again
on the imaginary axis at $\mu \approx 13.75$ and reappear as pairs of imaginary eigenvalues
of the opposite Krein signatures. This reversed instability bifurcation takes place in a complete agreement
with the necessary condition for the instability bifurcations.

In the large chemical potential limit, the eigenfrequencies of
the linearization at the excited state with $n = 2$ include the same
eigenfrequencies of the linearization at the ground state with $n = 0$ given by (\ref{tflimit}), see
the top right panel of Fig.~\ref{fig1}. In addition, two modes
with negative Krein signature appear due to the dynamics of the two
dark solitary waves on the ground state. One mode represents the in-phase
oscillation of the two dark solitons and it is
continued from the eigenvalue expanded by (\ref{pred4}) to the limit $\mu \to \infty$,
where it approaches $\omega_* = \sqrt{2}$. The other mode represents
the out-of-phase oscillation of the two dark solitons and it appears
from the complex pair (\ref{pred5}) which reappears back on the imaginary axis
for higher values of the chemical potential $\mu$.
Asymptotic approximation of the out-of-phase oscillation
in the limit $\mu \to \infty$ is reported in \cite{coles}.

This pattern continues for other excited states with $n \geq 3$.
The bottom panel on Fig.~\ref{fig1} shows the case $n = 3$.
For every $n \geq 3$, there are $n$ double eigenvalues with opposite
Krein signature at $\epsilon = 0$. If $\epsilon \neq 0$, the lowest
double eigenvalue does not lead to instability due to its linear degeneracy with the dipolar
symmetry mode. The remaining $n-1$ double eigenvalues may yield instability bifurcations
with complex eigenvalues. For large $\mu$, these eigenvalues reappear on
the imaginary axis after the reversed instability bifurcations in agreement
with the necessary condition for the instability bifurcation.
The $n$ eigenvalues of negative Krein signature
characterize $n$ dark solitons on the top of the ground state solution.
As such, they provide a rather lucid example of the nature and relevance
the negative Krein signature concept. Further details can be found
in~\cite{coles} for the large $\mu$ case and in~\cite{worldsci} for the small $\mu$ case.

\section{Krein signature for the nonlinear $\mathcal{PT}$-symmetric Schr\"{o}dinger equation}
\label{sec-PT}

Next, we consider the $\mathcal{PT}$-symmetric NLS equation (\ref{nls}) with the potential (\ref{pot-PT}).
Taking the nonlinear stationary states in the form $u(x,t) = e^{-i \mu t} \phi(x)$ with
$\mu \in \mathbb{R}$, we obtain the following differential equation
for the complex-valued $\phi$:
\begin{equation}
\mu \phi(x) = -\phi''(x) + ( x^2 + 2 i \gamma x ) \phi(x) + |\phi(x)|^2 \phi(x),
\label{NLSstat}
\end{equation}
where we have set $\Omega = 1$ again without loss of generality. We say that $\phi$
is a $\mathcal{PT}$-symmetric stationary state of the $\mathcal{PT}$-symmetric NLS equation
if $\phi$ satisfies the $\mathcal{PT}$-symmetry condition:
\begin{equation}
\label{PT-sym}
\phi(x) = \mathcal{PT} \phi(x) = \overline{\phi(-x)}, \quad x \in \mathbb{R}.
\end{equation}

In the linear (small-amplitude) limit, we can convert the linear spectral problem
to the quantum harmonic oscillator by using the complex variable $z = x + i \gamma$.
Then, the eigenvalues occur at $\mu_n = 1 + 2n +\gamma^2$, $n \in \mathbb{N}_0$
and the $\mathcal{PT}$-symmetric eigenfunctions are given by
\begin{equation}
\label{eigenvector-PT}
\varphi_n(x) = \frac{i^n}{\sqrt{2^n n! \sqrt{\pi}}} H_n(x+i\gamma) e^{-(x+i\gamma)^2/2}.
\end{equation}
Note that $\varphi_n$ in (\ref{eigenvector-PT}) satisfies
the $\mathcal{PT}$-symmetry condition (\ref{PT-sym}).
The eigenfunction $\varphi_n$ is normalized by the condition
\begin{equation}
\label{squared-norm}
\langle \varphi_n, \varphi_n \rangle_{\mathcal{PT}} = (-1)^n,
\end{equation}
where the modified inner product is used in the form
\begin{equation}
\langle \psi, \varphi \rangle_{\mathcal{PT}} := \int_{\mathbb{R}} \psi(x) \overline{\varphi(-x)} dx.
\label{Krein-PT-Linear}
\end{equation}
The inner product in the form (\ref{Krein-PT-Linear}) is used for all linear $\mathcal{PT}$-symmetric systems \cite{bender}
and the alternating sign of $\langle \varphi_n, \varphi_n \rangle_{\mathcal{PT}}$ is taken in \cite{yang}
as the Krein signature of the eigenvalue $\mu_n$, see discussion in Section 4.

By the same Crandall-Rabinowitz bifurcation theory \cite{CR}, each $\mathcal{PT}$-symmetric function $\varphi_n$
for a simple eigenvalue $\mu_n$ generates a branch of solutions, which can also be approximated by the same expansion
(\ref{expansion}). Bifurcations of such nonlinear stationary states in the $\mathcal{PT}$-symmetric systems
from simple real eigenvalues are considered in \cite{dohnal,PKT}, where it is proven that the
bifurcating branch of the stationary states satisfies the $\mathcal{PT}$-symmetry (\ref{PT-sym})
and the chemical potential $\mu$ is real (at least for small $\epsilon$).

The formal solvability condition for the correction terms $(\mu_n^{(2)},\varphi_n^{(3)})$
of the expansion (\ref{expansion}) yields
\begin{eqnarray}
\mu_n^{(2)} = \frac{\int_{\mathbb{R}} \varphi_n(x) |\varphi_n(x)|^2 \overline{\varphi_n(-x)} dx}{
\int_{\mathbb{R}} \varphi_n(x) \overline{\varphi_n(-x)} dx} = (-1)^n \int_{\mathbb{R}} \varphi_n(x) |\varphi_n(x)|^2 \overline{\varphi_n(-x)} dx.
\label{eqn2-PT}
\end{eqnarray}
Although it is obvious that $\mu_n^{(2)}$ is real, the sign of this quantity is less explicit than
in (\ref{eqn2}). At least for small $\gamma$, we know that $\mu_n^{(2)} > 0$ by continuity
of $\mu_n^{(2)}$ in $\gamma$. Continuation of branches of the nonlinear stationary states
in the limit $\mu \to \infty$ is a highly non-trivial problem (see \cite{zezyulin} for numerical results
and \cite{gallo} for partial analytical results on the ground state branch).

\begin{figure}[htb]
\begin{center}
\includegraphics[width=8cm]{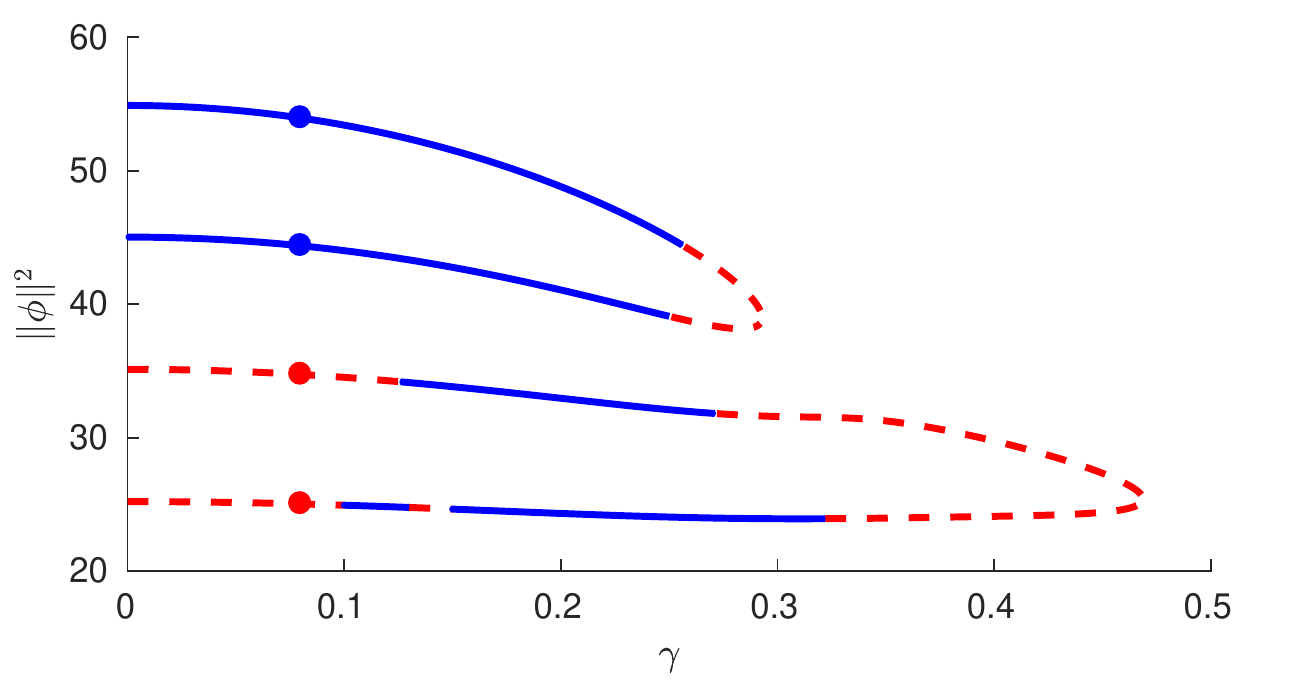}
\includegraphics[width=8cm]{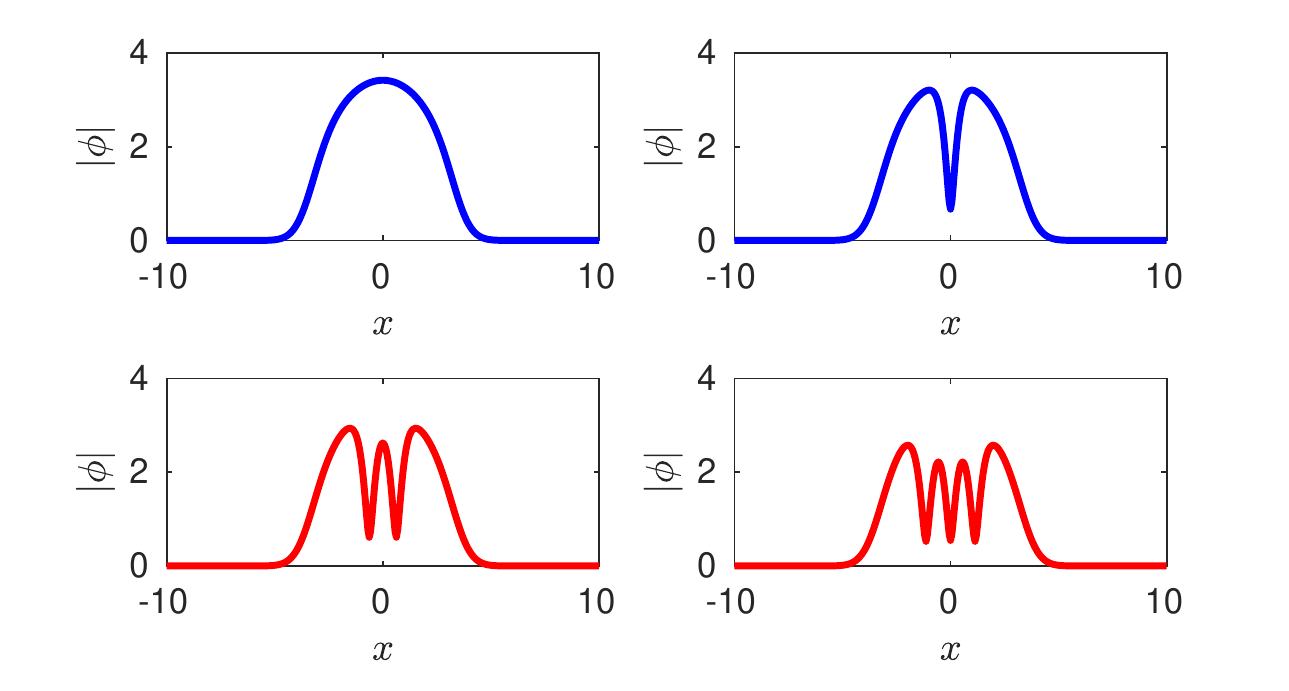}
\caption{Left: Power curves for branches of nonlinear states for $\mu=12$ and $\gamma > 0$.
Solid blue (dashed red) curves indicate stable (unstable) states.
Right: Sample profiles for nonlinear states that correspond to the points shown on the power curves,
from the top to the bottom branches.}
\label{figb}
\end{center}
\end{figure}

In our numerical experiments, we fix $\mu=12$ and continue in $\gamma$ first four branches from the
Hamiltonian case $\gamma=0$. The resulting continuations are shown on the left panel of Figure~\ref{figb}.
Branches with stable nonlinear states are shown by using blue solid curves and
branches with unstable states are shown in dashed red. The power curves represent the power of the mode:
$$
\|\phi\|^2 = \int_\mathbb{R} |\phi(x)|^2 dx.
$$
The right panel of Figure~\ref{figb} shows the mode profiles corresponding
to the points shown on the power branches on the left panel. Analyzing branches
reveals two saddle-node bifurcations: the first branch meets the second one at $\gamma\approx 0.292$,
whereas the third and fourth branches meet at $\gamma\approx 0.469$.
Profiles of the nonlinear states for the merging branches at the saddle-node bifurcation become
very similar, and after the bifurcation point both branches disappear.
Such bifurcations are typical in the defocusing case, whereas branches of nonlinear states
are extended for all $\gamma$ in the focusing case \cite{zezyulin}.

Linearizing the $\mathcal{PT}$-symmetric NLS equation with the same expansion
(\ref{eqn1}) yields the same spectral problem as in (\ref{spectrum}):
\begin{eqnarray}
\mathcal{L}(\gamma) v = -i \lambda \sigma_3 v,
\label{spectrum-PT}
\end{eqnarray}
with $\sigma_3 = {\rm diag}(1,-1)$, but $\mathcal{L}(\gamma)$ is no longer a self-adjoint linear operator.
The operator $\mathcal{L}(\gamma)$ is still defined in $L^2(\mathbb{R})$ with
the domain $H^2(\mathbb{R}) \cap L^{2,2}(\mathbb{R})$ and is now given by
\begin{equation}
\label{original}
\mathcal{L}(\gamma) = \left[\begin{array}{cc}
      -\partial^2_x + x^2 + 2 i \gamma x - \mu + 2|\phi(x)|^2 & \phi(x)^2 \\
      \overline{\phi(x)^2} & -\partial^2_x + x^2  - 2 i \gamma x - \mu + 2|\phi(x)|^2
\end{array}\right].
\end{equation}
This operator does not satisfy the Hamiltonian symmetry, $\sigma_3 \mathcal{L}(\gamma) \neq \mathcal{\bar{L}}(\gamma) \sigma_3$
but instead, it satisfies the $\mathcal{PT}$-symmetry $\mathcal{P} \mathcal{L}(\gamma) = \mathcal{\bar{L}}(\gamma) \mathcal{P}$.
In addition, it satisfies the symmetry $\sigma_1 \mathcal{L}(\gamma) = \mathcal{\bar{L}}(\gamma) \sigma_1$,
the same as in the Hamiltonian case. The quadruple symmetry of eigenvalues still exists due to these two symmetries.
Indeed, if $\lambda_0$ is an eigenvalue with the eigenvector $v_0 = (a,b)^T$,
then $-\bar\lambda_0$ is also an eigenvalue with the eigenvector
$\mathcal{PT} v_0$, that is $(\overline{a(-x)},\overline{b(-x)})^T$ for $x \in \mathbb{R}$, whereas
$\bar{\lambda}_0$ is another eigenvalue with the eigenvector $\sigma_1 \bar{v}_0 = (\bar{b},\bar{a})$.
Hence, eigenvalues of the $\mathcal{PT}$-symmetric spectral problem (\ref{spectrum-PT})
still occur either in real or purely imaginary pairs or as quadruplets in the complex plane.

Besides the spectral problem (\ref{spectrum-PT}), we also introduce the adjoint
spectral problem with the adjoint eigenvector denoted by $v^*$:
\begin{gather}
\mathcal{L}^*(\gamma) v^* = -i\lambda \sigma_3 v^*,
\label{Adj}
\end{gather}
where
\begin{gather}
\mathcal{L}^*(\gamma) = \left[\begin{array}{cc}
      -\partial^2_x  + x^2 - 2 i \gamma x - \mu + 2|\phi(x)|^2 & \phi(x)^2 \\
      \overline{\phi(x)^2} & -\partial^2_x + x^2 + 2 i \gamma x - \mu + 2|\phi(x)|^2
\end{array}\right].
\end{gather}
Unfortunately, the main limitations towards the Krein signature theory
in the $\mathcal{PT}$-symmetric case $\gamma \neq 0$ is that the adjoint eigenvector $v^*$
of the adjoint spectral problem (\ref{Adj}) cannot be
related to the eigenvector $v$ of the spectral problem (\ref{spectrum}) for the same eigenvalue $\lambda$.
Neither $\mathcal{L}^*(\gamma) = \mathcal{L}(\gamma)$ nor $\mathcal{L}^*(\gamma) = \mathcal{P} \mathcal{L}(\gamma) \mathcal{P}$
is true.

Let us now consider a simple isolated eigenvalue $\lambda_0 \in \mathbb{C} \backslash \{0\}$
of the spectral problems (\ref{spectrum-PT}) and (\ref{Adj})
with the eigenvector $v_0 \in H^2(\mathbb{R}) \cap L^{2,2}(\mathbb{R})$
and the adjoint eigenvector $v_0^* \in H^2(\mathbb{R}) \cap L^{2,2}(\mathbb{R})$, respectively.
If $\lambda_0\in i\mathbb{R}$, then there exists a choice for the eigenvectors $v_0$
and $v_0^*$ to satisfy the $\mathcal{PT}$-symmetry constraint:
\begin{equation}
\label{PT-sym-eigenvector}
v_0(x) = \overline{v_0(-x)}, \quad v_0^*(x) = \overline{v_0^*(-x)},  \quad  x \in \mathbb{R}.
\end{equation}
For each nonzero eigenvalue $\lambda_0 \in \mathbb{C}$ of the $\mathcal{PT}$-symmetric spectral problem (\ref{spectrum-PT})
with the eigenvector $v_0 = (a,b) \in H^2(\mathbb{R}) \cap L^{2,2}(\mathbb{R})$, we define
the Krein quantity $K(\lambda_0)$ as follows:
\begin{equation}
K(\lambda_0) := \langle \sigma_3 v_0, v_0^* \rangle
    = \int_{\mathbb{R}} \left(a(x) \overline{a^*(x)} - b(x)\overline{b^*(x)} \right) dx.
\label{pt-krein}
\end{equation}

If $\gamma = 0$, then $\mathcal{L}^*(0) = \mathcal{L}(0)$ and
the adjoint spectral problem (\ref{Adj}) becomes equivalent to the
spectral problem (\ref{spectrum-PT}). Therefore, the adjoint eigenvector $v_0^*$
can be related to the eigenvector $v_0$ by $v_0^* = v_0$. In this Hamiltonian case, the definition (\ref{pt-krein}) represents the
integral in the right-hand-side of the definition (\ref{eqn2b}).
The signs of $K(\lambda_0)$ defined for $\gamma = 0$ by (\ref{eqn2b}) and
$K(\lambda_0)$ defined for $\gamma \in \mathbb{R}$ by (\ref{pt-krein}) are
the same if $-i \lambda_0 > 0$ and $\gamma = 0$.

If $\gamma \neq 0$, the adjoint eigenvector $v_0^*$ satisfying the $\mathcal{PT}$-symmetry condition (\ref{PT-sym-eigenvector})
is defined up to an arbitrary sign. As a result, the Krein quantity $K(\lambda_0)$ in (\ref{pt-krein}) is defined up to
the sign change. In the continuation of the NLS equation (\ref{nls})
with respect to the parameter $\gamma$ from the Hamiltonian case $\gamma = 0$,
the sign of the Krein quantity $K(\lambda_0)$ in (\ref{pt-krein}) is chosen so that
it matches the sign of $K(\lambda_0)$ in (\ref{eqn2b}) for every $-i \lambda_0 > 0$ and $\gamma = 0$,
hence we choose $v_0^* = v_0$ at $\gamma = 0$. With this convention on the normalization of the adjoint
eigenvectors, the eigenvector $v_0$, the adjoint eigenvector $v_0^*$, and
the Krein quantity $K(\lambda_0)$ are extended continuously with respect to the parameter $\gamma$.

Let us verify the three main properties of the Krein quantity $K(\lambda_0)$ defined by (\ref{pt-krein}).
\begin{enumerate}
\item If $f$ and $g$ satisfy the $\mathcal{PT}$-symmetry condition (\ref{PT-sym-eigenvector}),
then the standard inner product $\langle f, g\rangle$ is real-valued. Indeed, this follows from
\begin{eqnarray*}
        \langle f,g\rangle = \int_{\mathbb{R}} f(x)\overline{g(x)} dx & = &
        \int_0^{+\infty} \bigl( f(x)\overline{g(x)} + f(-x)\overline{g(-x)}\bigr)dx \\
        & = & \int_0^{+\infty} \bigl( f(x)\overline{g(x)} + \overline{f(x)}g(x) \bigr) dx.
\end{eqnarray*}
By~\eqref{PT-sym-eigenvector}, $v_0$ and $v_0^*$ are $\mathcal{PT}$-symmetric if $\lambda_0 \in i \mathbb{R}$,
hence $K(\lambda_0)$ is real if $\lambda_0 \in i \mathbb{R}$.

\item
Let us write the spectral problem ~\eqref{original} for the generalized eigenvector $v_g$:
\begin{equation}
\label{pt-non-homog}
           (\mathcal{L}(\gamma) + i \lambda_0 \sigma_3 )v_g
           = \sigma_3 v_0.
\end{equation}
If $\lambda_0 \in i\mathbb{R}\backslash \{0\}$, then $v_0^*$ is in the kernel of the adjoint operator
$(\mathcal{L}(\gamma) + i\lambda_0\sigma_3)^*$, and
Fredholm solvability condition of the above equation
is $\langle \sigma_3 v_0,v_0^*\rangle = 0$. If $K(\lambda_0) = 0$, then there exists a solution
to the nonhomogeneous equation (\ref{pt-non-homog}), so that $\lambda_0$ is not simple.
Hence, $K(\lambda_0) \neq 0$.

\item Taking inner products of the spectral problems \eqref{spectrum-PT} and \eqref{Adj} with the corresponding
eigenvectors yields
\begin{equation*}
   \left\langle \mathcal{L} v_0, v_0^* \right\rangle = -i\lambda_0
   \left\langle\sigma_3 v_0, v_0^*\right\rangle
\end{equation*}
and
\begin{equation*}
   \left\langle v_0, \mathcal{L}^* v_0^* \right\rangle
           = i\overline{\lambda}_0
   \left\langle v_0,   \sigma_3 v_0^*   \right\rangle,
\end{equation*}
hence
\begin{equation*}
i (\lambda_0 + \overline{\lambda}_0) K(\lambda_0) = 0.
\end{equation*}
If $\lambda_0 \notin i\mathbb{R}$,
then $\lambda_0 + \overline{\lambda}_0 \neq 0$ and $K(\lambda_0) = 0$.
\end{enumerate}

Let us now illustrate how the Krein signatures can be used to predict instability bifurcations
from multiple neutrally stable eigenvalues of the spectral problem (\ref{spectrum-PT}).
Recall that the eigenvalue is called {\em semi-simple} if algebraic and geometric multiplicities coincide
and {\em defective} if algebraic multiplicity exceeds geometric multiplicity.
In Section 2, we continued a semi-simple double eigenvalue with respect to parameter $\epsilon$.
Here we continue a defective double eigenvalue with respect to parameter $\gamma$.

Let $\gamma_0$ denote the bifurcation point when two neutrally stable eigenvalues coalesce:
$\lambda_0 = \lambda_0' \in i \mathbb{R} \backslash \{0\}$.
Near $\gamma = \gamma_0$, we expand the linear non-self-adjoint operator $\mathcal{L}(\gamma)$
in (\ref{original}) as follows:
\begin{equation}
\mathcal{L}(\gamma) = \mathcal{L}_0 + (\gamma - \gamma_0) \mathcal{L}_1 + \dots,
\label{operator:series}
\end{equation}
where
\begin{equation}
\mathcal{L}_1 =
\left[\begin{array}{cc}
        2 ix  + 2\partial_\gamma |\phi(x)|^2 |_{\gamma = \gamma_0} & \partial_\gamma \phi^2(x) |_{\gamma = \gamma_0} \\
        \partial_\gamma \overline{\phi^2(x)} |_{\gamma = \gamma_0} & -2ix + 2 \partial_\gamma |\phi(x)|^2 |_{\gamma = \gamma_0}
\end{array}\right],
\label{operator:lprime}
\end{equation}
and $\partial_\gamma$ denotes a partial derivative with respect to the parameter $\gamma$.
We assume that there exists a defective double eigenvalue $\lambda_0 \in i \mathbb{R} \backslash \{0\}$
of the spectral problems (\ref{spectrum-PT}) and (\ref{Adj})
with the eigenvector $v_0$, the generalized eigenvector $v_g$,
the adjoint eigenvector $v_0^*$, and the adjoint generalized eigenvector $v_g^*$, respectively.
We will show that under the following non-degeneracy condition
\begin{equation}
\label{non-degeneracy}
\langle \mathcal{L}_1 v_0, v_0^*\rangle \ne 0,
\end{equation}
the necessary condition for instability bifurcation is satisfied in the continuations
with respect to the parameter $\gamma$.
Thanks to the decomposition~\eqref{operator:series}, we are looking for an eigenvalue $\lambda(\gamma)$
of the perturbed spectral problem
\begin{equation}
\left( \mathcal{L}_0
	+ (\gamma-\gamma_0) \mathcal{L}_1 + \ldots \right) v(\gamma) =
-i\lambda(\gamma) \sigma_3 v(\gamma),
\label{original:perturbed}
\end{equation}
such that $\lambda(\gamma) \to \lambda_0$ as $\gamma \to \gamma_0$.
Since $\lambda_0$ is a defective eigenvalue of geometric multiplicity {\em one}
and algebraic multiplicity {\em two}, we apply Puiseux expansions~\cite{Knopp}:
\begin{equation}
\left\{ \begin{array}{l}
\lambda(\gamma) = \lambda_0 + (\gamma-\gamma_0)^{1/2} \lambda_g
                    + (\gamma-\gamma_0) \tilde{\lambda} + \ldots, \\
v(\gamma) = v_0 -i (\gamma-\gamma_0)^{1/2} \lambda_g v_g
                    + (\gamma-\gamma_0) v_1 + \ldots,
                    \end{array} \right.
                    \label{Puiseux}
\end{equation}
where $\lambda_g$, $\tilde{\lambda}$, and $v_1$ are correction terms. To define $v_1$ uniquely, we add the orthogonality
condition $\langle \sigma_3 v_1, v_0^* \rangle = \langle \sigma_3 v_1, v_g^* \rangle = 0$.
The coefficient $-i\lambda_g$ comes in front of $v_g$ thanks to the nonhomogeneous equation
(\ref{pt-non-homog}) arising at the order of $(\gamma-\gamma_0)^{1/2}$ from
the perturbed spectral problem \eqref{original:perturbed}.

Plugging (\ref{Puiseux}) into~\eqref{original:perturbed} yields at the order of
$(\gamma-\gamma_0)$:
\begin{equation}
\label{another-non-homog}
\left( \mathcal{L}_0 + i\lambda_0 \sigma_3 \right) v_1
    = -\mathcal{L}_1 v_0 - \lambda_g^2 \sigma_3 v_g - i \tilde{\lambda} \sigma_3 v_0.
\end{equation}
Fredholm solvability condition is satisfied if the right-hand side of the nonhomogeneous equation (\ref{another-non-homog})
is orthogonal to the kernel of adjoint operator $(\mathcal{L}_0 + i\lambda_0\sigma_3)^*$ spanned by $v_0^*$. This orthogonality condition
yields the constraint:
\begin{equation}
\label{proj-eq}
    \langle -\mathcal{L}_1 v_0 - \lambda_g^2 \sigma_3 v_g - i\tilde{\lambda} \sigma_3 v_0,
    v_0^* \rangle = 0.
\end{equation}
Since $K(\lambda_0) = 0$ for the defective eigenvalue $\lambda_0\in i\mathbb{R}$,
$\tilde{\lambda}$ is not determined by equation (\ref{proj-eq}). On the other hand,
$\lambda_g$ is defined by equation (\ref{proj-eq}), which can be rewritten as follows:
\begin{equation}
    (-i\lambda_g)^2 = \frac{\langle \mathcal{L}_1 v_0,v_0^*\rangle}
                      {\langle \sigma_3 v_g, v_0^*\rangle}.
                      \label{eq:omegag}
\end{equation}
The denominator of (\ref{eq:omegag}) is nonzero because of the following argument.
If $\lambda_0$ is a double eigenvalue, then the solution of the nonhomogeneous equation
\begin{equation*}
           (\mathcal{L}_0 + i \lambda_0 \sigma_3 ) \tilde{v}_g
           = \sigma_3 v_g,
\end{equation*}
does not exist in $L^2(\mathbb{R})$. Hence $\langle \sigma_3 v_g, v_0^*\rangle \neq 0$.
Since $v_0$, $v_0^*$, $v_g$, and $\mathcal{L}_1$ satisfy the
$\mathcal{PT}$-conditions (\ref{PT-sym}) and (\ref{PT-sym-eigenvector}), both
the nominator and the denominator of (\ref{eq:omegag}) are real-valued.
By the assumption (\ref{non-degeneracy}),
the numerator of (\ref{eq:omegag}) is nonzero. Thus,
$(-i\lambda_g)^2$ is either positive or negative.

Let us assume that $(-i\lambda_g)^2 > 0$ without loss of generality.
If $\gamma > \gamma_0$, then $i(\gamma-\gamma_0)^{1/2} \lambda_g \in \mathbb{R}$ and
we obtain the following expansions for the two simple purely imaginary eigenvalues
$\lambda_1$ and $\lambda_2$ given by
\begin{align}
    \lambda_1 &= \lambda_0 + (\gamma-\gamma_0)^{1/2} \lambda_g
        + \ldots, \\
    \lambda_2 &= \lambda_0 - (\gamma-\gamma_0)^{1/2} \lambda_g
        + \ldots
\end{align}
The corresponding eigenvectors are expanded by
\begin{align}
    v_1(\gamma) &= v_0 - i(\gamma-\gamma_0)^{1/2} \lambda_g v_g
        + \ldots, \\
    v_2(\gamma) &= v_0 + i(\gamma-\gamma_0)^{1/2} \lambda_g v_g
        + \ldots,
\end{align}
whereas the adjoint eigenvectors for the same eigenvalues are expanded by
\begin{align}
    v_1^*(\gamma) &= v_0^* - i(\gamma-\gamma_0)^{1/2} \lambda_g v_g^*
        + \ldots, \\
    v_2^*(\gamma) &= v_0^* + i(\gamma-\gamma_0)^{1/2} \lambda_g v_g^*
        + \ldots
\end{align}
The leading order of Krein quantitites for eigenvalues $\lambda_1$ and $\lambda_2$
is given by
\begin{align}
    K(\lambda_1) &= \langle \sigma_3 v_1(\gamma),v_1^*(\gamma)\rangle
    = -i(\gamma-\gamma_0)^{1/2} \lambda_g
    \langle \sigma_3 v_g,v_0^*\rangle + i\overline{(\gamma-\gamma_0)^{1/2} \lambda_g}
        \langle \sigma_3 v_0,v_g^*\rangle + \ldots,
        \label{krein:1} \\
        K(\lambda_2) &= \langle \sigma_3 v_2(\gamma),v_2^*(\gamma)\rangle =
        +i(\gamma-\gamma_0)^{1/2} \lambda_g \langle \sigma_3 v_g,v_0^*\rangle
    - i\overline{(\gamma-\gamma_0)^{1/2} \lambda_g} \langle \sigma_3 v_0,v_g^*\rangle
    + \ldots
    \label{krein:2}
\end{align}
Since
$$
\langle \sigma_3 v_g,v_0^*\rangle = \langle v_g, \sigma_3 v_0^* \rangle =
\langle v_g, (\mathcal{L}_0 + i \lambda_0 \sigma_3 )^* v_g^* \rangle =
\langle (\mathcal{L}_0 + i \lambda_0 \sigma_3 ) v_g, v_g^* \rangle =
\langle \sigma_3 v_0,v_g^*\rangle
$$
the two expansions for $K(\lambda_1)$ and $K(\lambda_2)$ can be rewritten
in the case of $i(\gamma-\gamma_0)^{1/2} \lambda_g \in \mathbb{R}$ as
\begin{align*}
    K(\lambda_1) &= -2i(\gamma-\gamma_0)^{1/2} \lambda_g \langle \sigma_3 v_g, v_0^*\rangle
            + \ldots, \\
            K(\lambda_2) &= 2i(\gamma-\gamma_0)^{1/2} \lambda_g \langle \sigma_3 v_g,v_0^*\rangle
                + \ldots
\end{align*}
Since $\langle \sigma_3 v_0, v_g^* \rangle \neq 0$, $K(\lambda_1)$ has the opposite sign to $K(\lambda_2)$.

If $\gamma < \gamma_0$, then $i(\gamma-\gamma_0)^{1/2} \lambda_g \in i \mathbb{R}$,
so that $\lambda_1, \lambda_2 \notin i \mathbb{R}$, whereas $K(\lambda_1) = K(\lambda_2) = 0$.
Thus, the necessary condition for the instability bifurcation holds under the nondegeneracy assumption
(\ref{non-degeneracy}).

Note in passing that if the non-degeneracy assumption (\ref{non-degeneracy}) is not satisfied, then
$\lambda_g = 0$ follows from (\ref{eq:omegag})
and the perturbation theory must be extended to the next order
with a characteristic equation to be derived for the correction term $\tilde{\lambda}$. In this case,
the double defective eigenvalue $\lambda_0\in i\mathbb{R}$ may split safely along $i \mathbb{R}$
both for $\gamma > \gamma_0$ and $\gamma < \gamma_0$.

\begin{figure}
\begin{center}
\includegraphics[width=8cm]{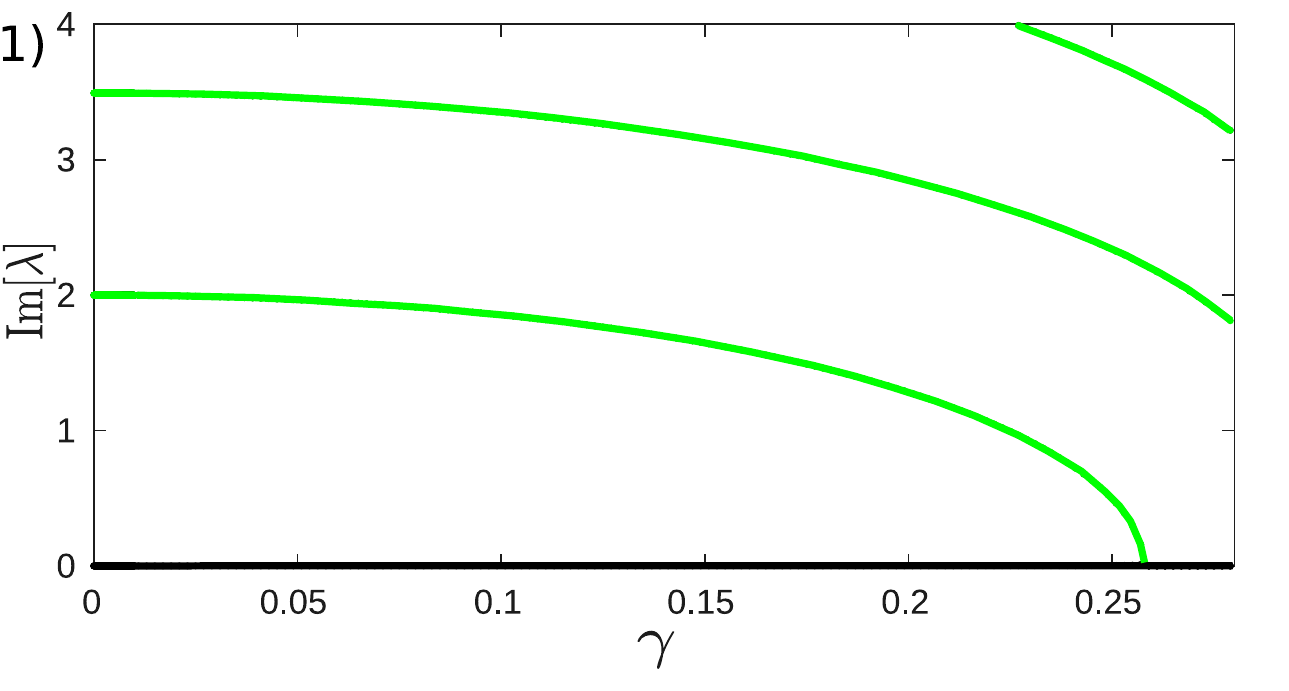}
\includegraphics[width=8cm]{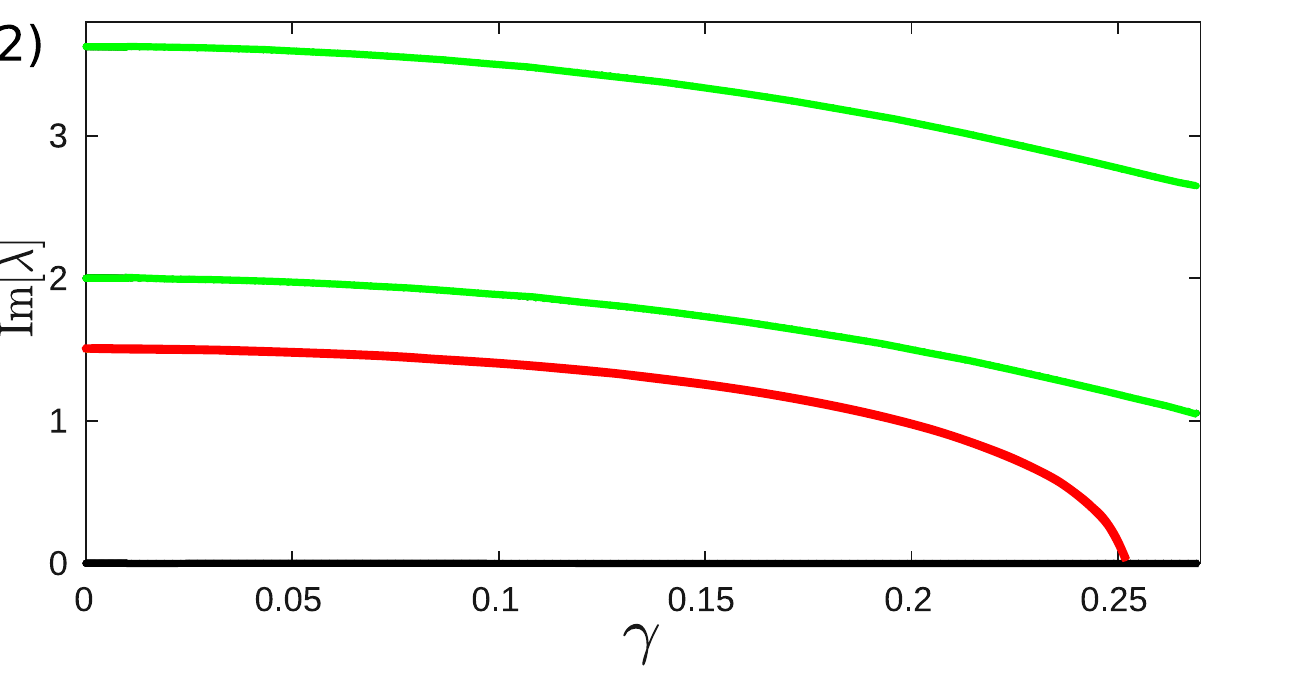}
\caption{Purely imaginary eigenvalues $\lambda$ of the $\mathcal{PT}$-symmetric problem (\ref{spectrum-PT})
for the first two stationary states with $\mu=12$. Eigenvalues of negative (positive) Krein signature
are shown in red (green), complex eigenvalues are shown in black.}
\label{fig2}
\end{center}
\end{figure}

\begin{figure}
\begin{center}
\includegraphics[width=8cm]{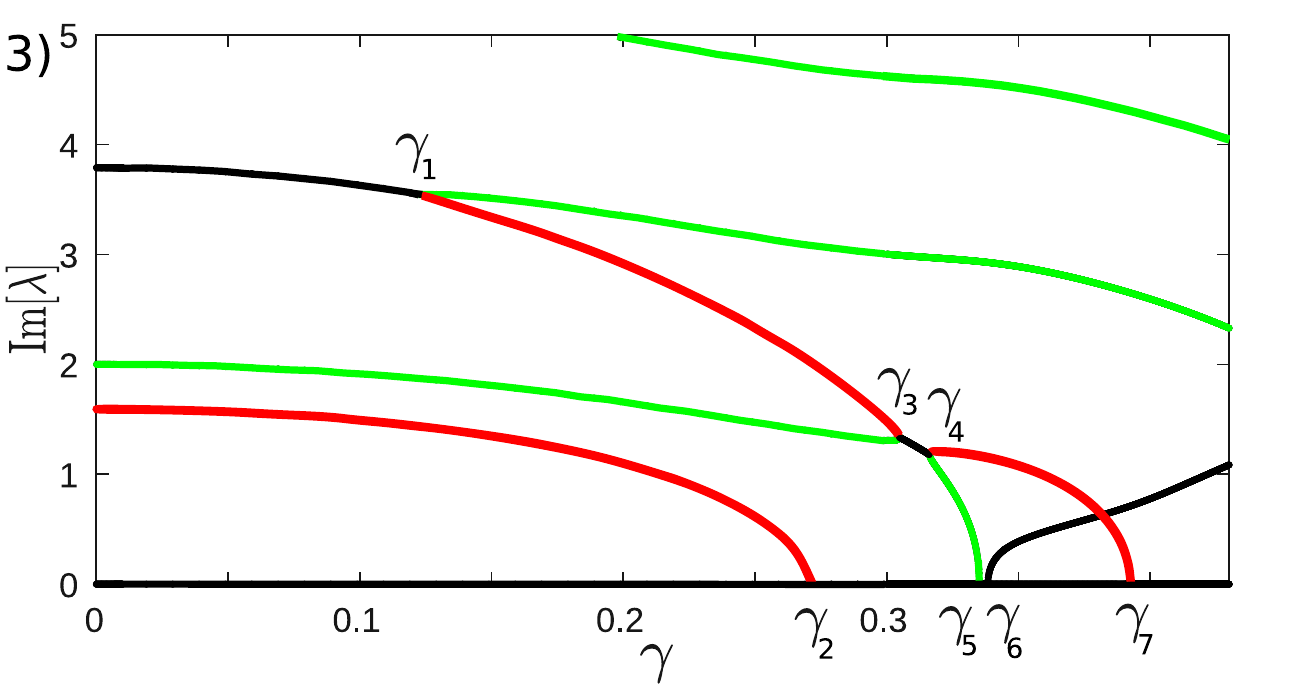}
\includegraphics[width=8cm]{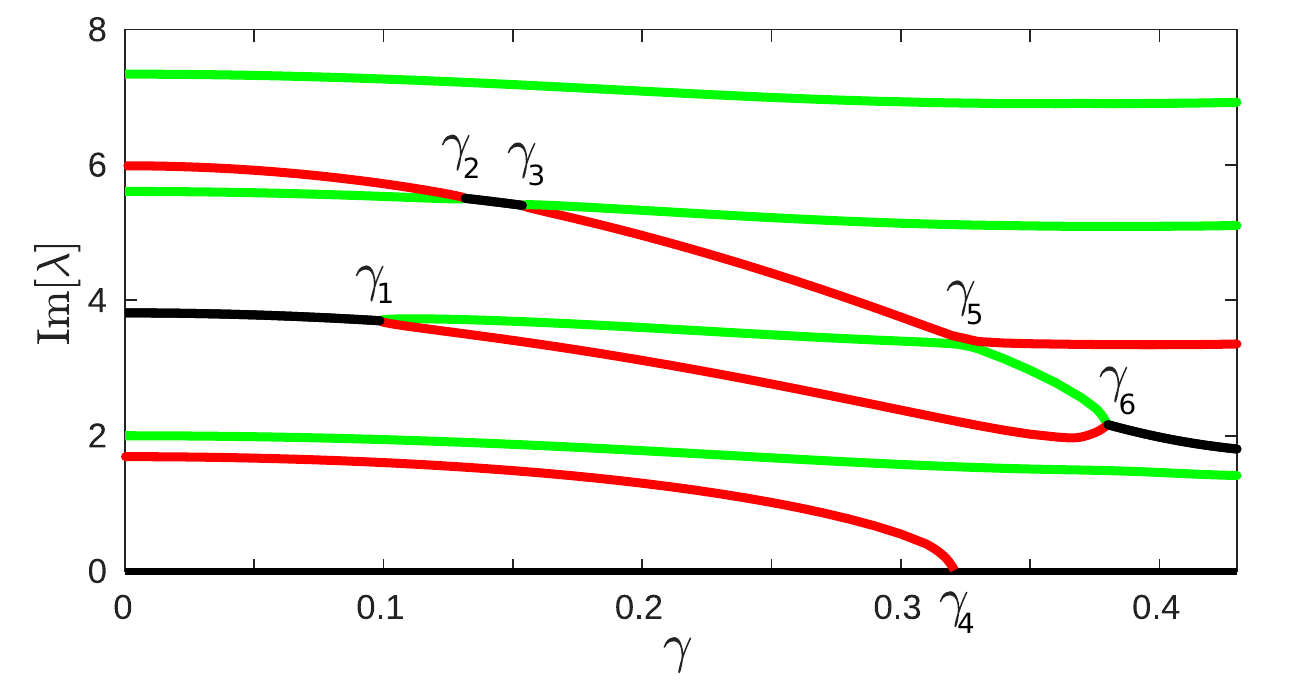}
\includegraphics[width=8cm]{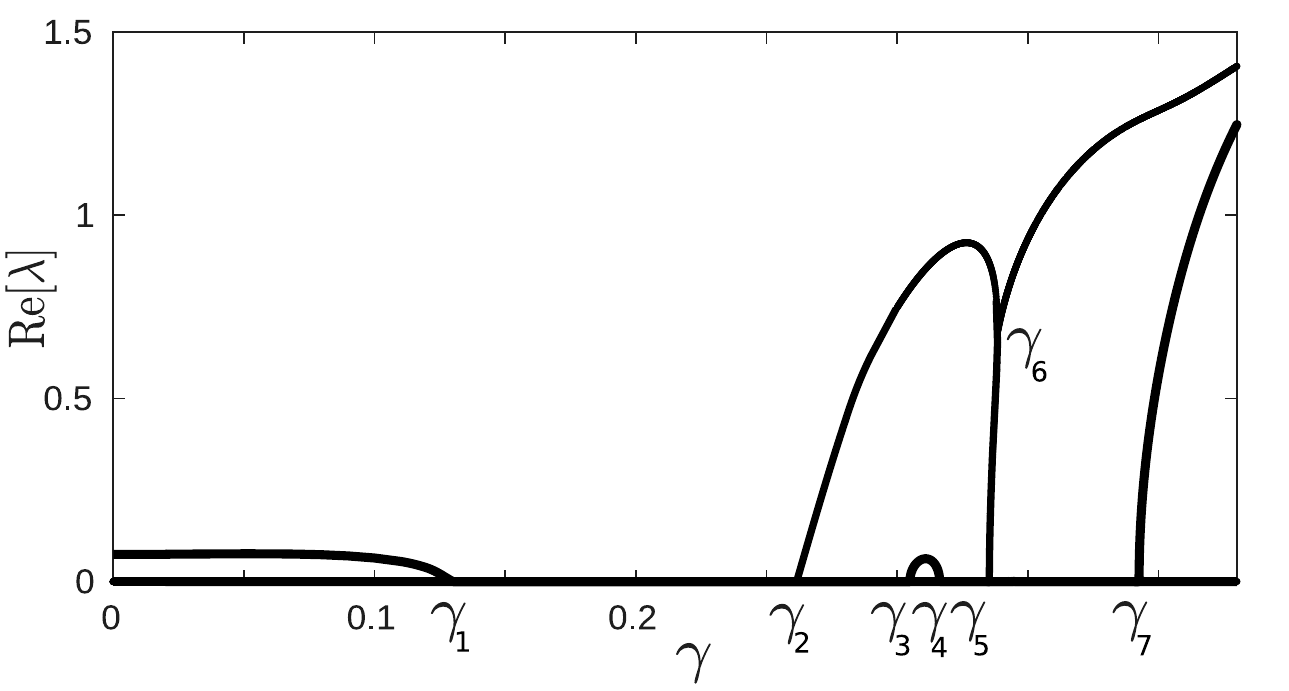}
\includegraphics[width=8cm]{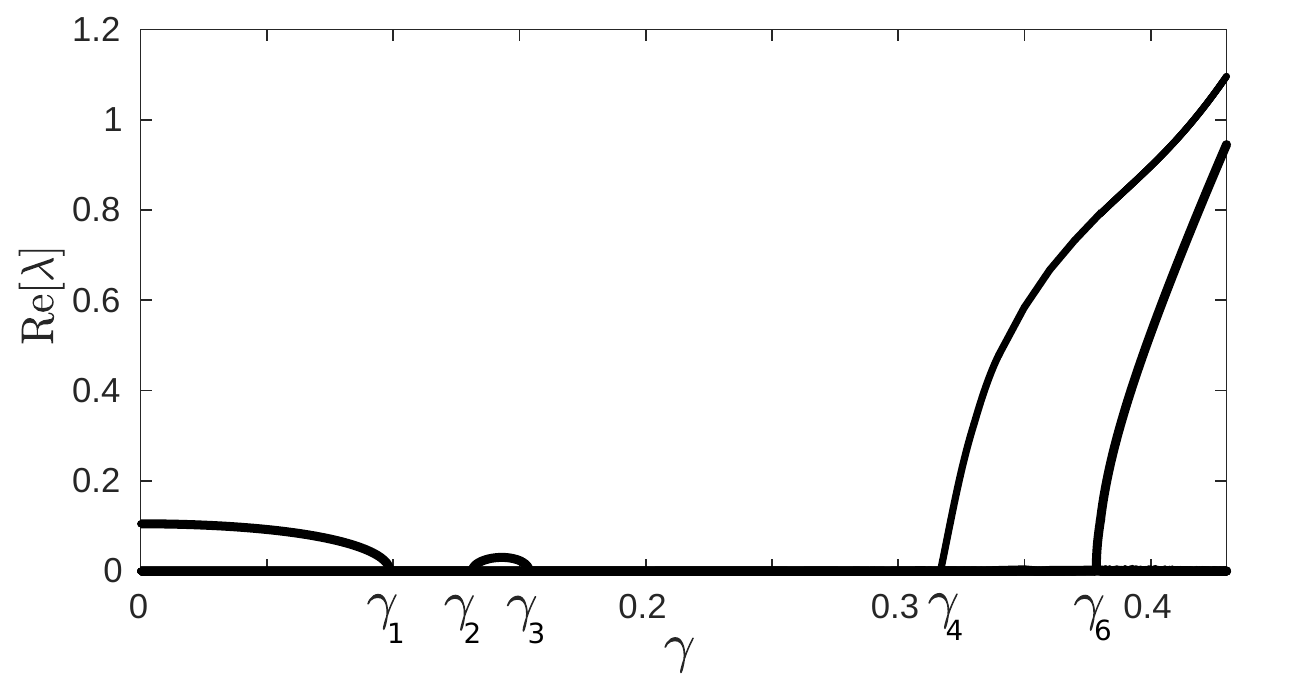}
\caption{The same as Figure \ref{fig2} but for the third (left) and fourth (right) branches
of the stationary states. Top panels show imaginary parts and the bottom panels show real parts
of the eigenvalues $\lambda$.}
\label{fig2-extra}
\end{center}
\end{figure}

Figures~\ref{fig2} and \ref{fig2-extra} show eigenvalues of the $\mathcal{PT}$-symmetric spectral problem
(\ref{spectrum-PT}) for the first four branches of the nonlinear stationary states with $\mu = 12$ shown
on Figure \ref{figb}.

Figure~\ref{fig2} shows that the first
branch is stable until $\gamma\approx 0.27$, whereas the second branch is stable until $\gamma\approx 0.25$.
For the first branch (left panel), eigenvalues of the positive Krein signature coalesce at the origin,
whereas for the second branch (right panel), eigenvalues of the negative Krein signature coalesce
at the origin. The instability of the first branch is unusual, since it plays the role of the
`ground state' in analogy to Hamiltonian case. Nonetheless, this is no surprise since similar behavior was
observed in~\cite{zezyulin}, where the first two branches lost their stability very close to each other.

Figure~\ref{fig2-extra} (left panels) shows seven bifurcations among eigenvalues of the third branch
of the stationary states that occur
at $\gamma_1 \approx 0.126$, $\gamma_2 \approx 0.271$, $\gamma_3 \approx 0.304$,
$\gamma_4 \approx 0.316$, $\gamma_5 \approx 0.335$, $\gamma_6 \approx 0.338$, and $\gamma_7 \approx 0.393$.
Bifurcations at $\gamma_2$, $\gamma_5$, and $\gamma_7$ occur when eigenvalues on the imaginary axis coalesce at the origin,
resulting in pair of eigenvalues on the real axis. The necessary condition for instability bifurcations
is developed for $\lambda_0 \in i \mathbb{R} \backslash \{0\}$ and it is not applicable if $\lambda_0 = 0$.
The bifurcation at $\gamma_6$ occurs when real eigenvalues formed after bifurcations at
$\gamma_2$ and $\gamma_5$ coalesce and transform into a quadruplet of complex eigenvalues.

At $\gamma_1$, complex quadruplets continued from the case $\gamma = 0$ coalesce and bifurcate into the
imaginary eigenvalues with opposite Krein signatures, which provides an excellent example
for the necessary condition of the reverse instability bifurcation. At $\gamma_3$ and $\gamma_4$,
we have more examples of the instability bifurcation and the reverse instability bifurcation,
in which the two eigenvalues before $\gamma_3$ and after $\gamma_4$
on the imaginary axis have opposite Krein signatures.

Figure~\ref{fig2-extra}~(right panels) shows six bifurcations among eigenvalues of the fourth branch
of the stationary states at $\gamma_1 \approx 0.099$, $\gamma_2 \approx 0.131$, $\gamma_3 \approx 0.154$,
$\gamma_4 \approx 0.322$, $\gamma_5\approx 0.326$ and $\gamma_6 \approx 0.380$.
The bifurcation at $\gamma_1$ is similar to the one for the third branch:
a complex pair of eigenvalues
coming from the Hamiltonian case coalesces on the imaginary axis and splits along the imaginary axis
into two eigenvalues with opposite Krein signatures moving away from each other.
Bifurcations at $\gamma_2$ and $\gamma_3$ occur when
two imaginary eigenvalues with opposite Krein signatures continued from $\gamma = 0$
coalesce and bifurcate off into the complex plane at $\gamma_2$, after which
the complex eigenvalues coalesce again on the imaginary axis at $\gamma_3$ and emerge as a pair of
purely imaginary eigenvalues with opposite Krein signatures.

\begin{figure}
\begin{center}
\includegraphics[width=8cm]{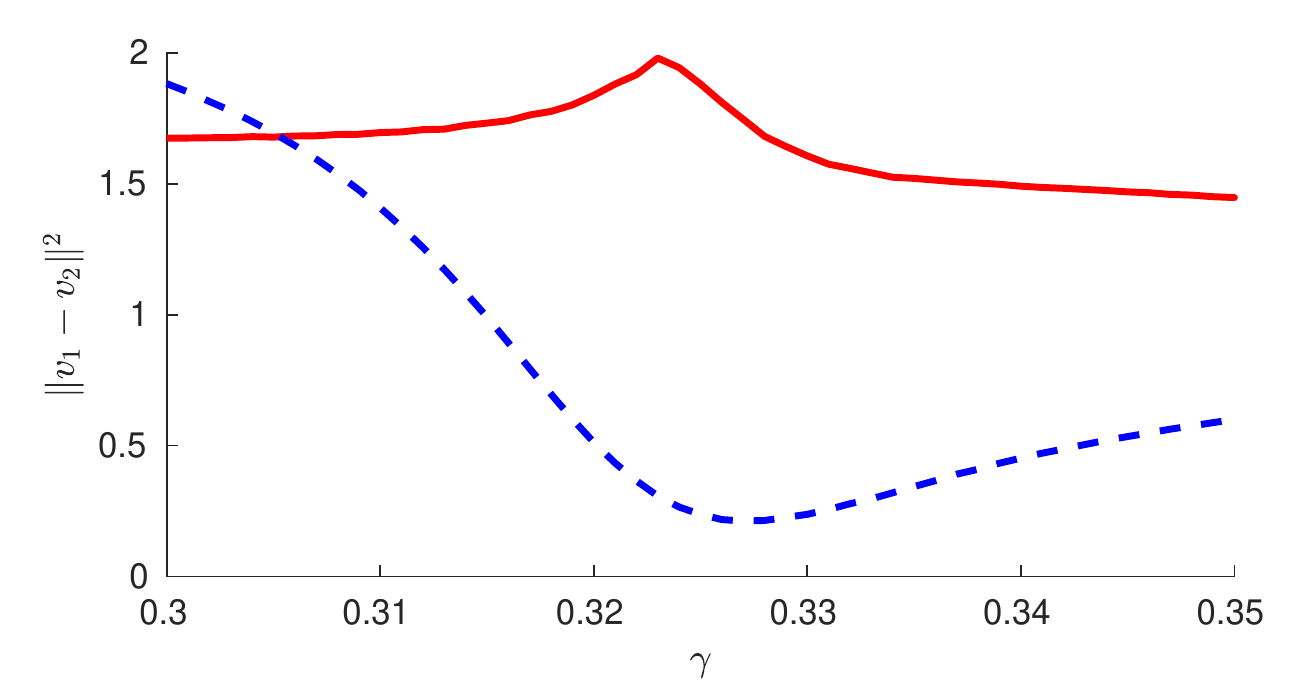}
\caption{Distance between eigenvectors and adjoint eigenvectors
  for the fourth branch in Figure~\ref{fig2-extra} near $\gamma_5$.
  The eigenvector distance
 is depicted by red solid curve. The adjoint eigenvector distance is depicted
by dotted blue curve.}
\label{fig3}
\end{center}
\end{figure}

At $\gamma_4$, a pair of purely imaginary eigenvalues of negative Krein signature coalesces at the origin
and they bifurcate into real eigenvalues. At $\gamma_5$, the purely imaginary eigenvalues nearly coalesce, but the
numerical results are somewhat inconclusive. Figure~\ref{fig3} shows the squared norm of the difference
of eigenvectors for the corresponding eigenvalues. As we can see, the difference between eigenvectors
does not vanish, which rules out the possibility
of bifurcation point due to a double defective eigenvalue.

Finally, bifurcation at $\gamma_6$ shows coalescence of two eigenvalues with opposite Krein signatures
after which they bifurcate into a complex quadruplet. Bifurcations at $\gamma_1$, $\gamma_2$,
$\gamma_3$, and $\gamma_6$ agree with the necessary condition for the instability bifurcation.

Stability of nonlinear stationary states of the $\mathcal{PT}$-symmetric NLS equation ~\eqref{NLSstat} was studied
numerically in~\cite{zezyulin} for fixed values of $\gamma$.
The recent study in~\cite{ChernPel} was applied to a modified potential $V$ where the imaginary part
of $V$ had a Gaussian decay; see also the earlier
study of~\cite{achilleos}. The instability bifurcations were found
to be very similar to the present study.
In addition to the bifurcations visible on Figure \ref{fig2-extra}, there was also the case when
two eigenvalues with  opposite signatures coalesce into a defective eigenvalue but not bifurcating
into the complex plane. This may happen when the non-degeneracy condition~\eqref{non-degeneracy} is not satisfied,
so that the two eigenvalues of opposite Krein signature can pass each other on the imaginary axis
without generating complex quadruplets.

\section{Krein signature for the linear $\mathcal{PT}$-symmetric Schr\"{o}dinger equation}
\label{sec-lin-PT}

Here we discuss the concept of Krein signature in the linear $\mathcal{PT}$-symmetric Schr\"{o}dinger equation
introduced in~\cite{yang}. For the potential (\ref{pot-PT}) with $\Omega = 1$,
we can write the linear spectral problem in the form:
\begin{equation}
\mu \psi(x) = -\psi''(x) + x^2 \psi(x) + 2 i \gamma x \psi(x),
\label{PTev}
\end{equation}
which is related to the non-self-adjoint $\mathcal{PT}$-symmetric Schr\"{o}dinger operator
$\mathcal{H} = -\partial^2_x + x^2 + 2 i \gamma x$ defined on the domain $H^2(\mathbb{R}) \cap L^{2,2}(\mathbb{R})$
in $L^2(\mathbb{R})$. The adjoint operator $\mathcal{H}^* = -\partial^2_x + x^2 - 2 i\gamma x$
satisfies $\mathcal{H}^* = \mathcal{P} \mathcal{H} \mathcal{P}$, where
$\mathcal{P}$ is the parity operator. Because of this relation,
if $\psi_0$ is an eigenfunction of $\mathcal{H}$ for the eigenvalue $\mu_0$,
then $\psi_0^* = \mathcal{P} \psi_0$ is an eigenfunction of $\mathcal{H}^*$
for the same eigenvalue $\mu_0$. By using the relation $\psi_0^*(x) = \psi_0(-x)$,
the Krein quantity of the eigenvalue $\mu_0$ in the spectral problem (\ref{PTev})
can be defined by the inner product in (\ref{Krein-PT-Linear}):
\begin{equation}
K(\mu_0) := \langle \psi_0, \psi_0 \rangle_{\mathcal{PT}} =
\langle \psi_0, \psi_0^* \rangle = \int_{\mathbb{R}} \psi_0(x) \overline{\psi_0(-x)} dx.
\label{lin-krein}
\end{equation}
This definition was used in \cite{yang} to verify the three properties of the
Krein quantity and the necessary condition for instability bifurcation.

The spectral problem (\ref{PTev}) can be written in the Hamiltonian form (\ref{lin-Ham}),
or explicitly,
\begin{equation}
i \mathcal{P} (\mathcal{P} \mathcal{H})\psi = i\mu \psi,
\label{PTev-pre-Ham}
\end{equation}
where $\mathcal{L} = \mathcal{P} \mathcal{H}$ is self-adjoint, $J = i \mathcal{P}$ is skew-adjoint and
invertible, and $\lambda = i \mu$ is a new eigenvalue.
By using the definition (\ref{krein-intro}) of the Krein quantity for the Hamiltonian
spectral problem (\ref{lin-Ham}), we obtain
\begin{equation}
\label{lin-krein-tilde}
\tilde{K}(\mu_0) = \langle \mathcal{P} \mathcal{H} \psi_0, \psi_0 \rangle = \mu_0 \langle \psi_0, \mathcal{P} \psi_0 \rangle =
\mu_0 K(\mu_0),
\end{equation}
which is only different from the definition~\eqref{lin-krein} by the factor $\mu_0$.
However, $\mu_0 > 0$ since the spectral problem (\ref{PTev}) admits only positive eigenvalues.
Thus, the Krein signature introduced in (\ref{lin-krein})
coincides with the Krein signature introduced in (\ref{lin-krein-tilde}).

The only difference between the Hamiltonian spectral problem (\ref{spectrum})
for the linearized NLS equation  and the spectral problem (\ref{PTev-pre-Ham})
for the linear Schr\"{o}dinger equation is that the eigenvalues $\lambda$
of the spectral problem (\ref{spectrum}) on the imaginary axis occur in pairs
thanks to the symmetry $\sigma_1 \mathcal{L} = \mathcal{\bar{L}} \sigma_1$, whereas
the eigenvalues $\lambda = i \mu$ of the spectral problem (\ref{PTev-pre-Ham}) are located
on the positive imaginary axis.

In the limit $\gamma \to 0$, eigenfunctions of the Schr\"{o}dinger operator
$\mathcal{H}_0 = -\partial^2_x + x^2$ for the quantum harmonic oscillator are either
even or odd. Eigenvalues $\mu_{2N} = 4N+1$, $N \in \mathbb{N}_0$
with even eigenfunctions have positive Krein signature in
(\ref{lin-krein}), whereas eigenvalues $\mu_{2N-1} = 4N-1$, $N \in \mathbb{N}$
with odd eigenfunctions
have negative Krein signature. This seems to be surprising at first glance,
since all eigenvalues are strictly positive and the operator $\mathcal{H}_0$
is self-adjoint in $L^2(\mathbb{R})$.

It is more natural in the Hamiltonian case $\gamma = 0$
to define the Krein quantity of an eigenvalue $\mu_0$ by
\begin{equation}
\label{lin-krein-last}
K_H(\mu_0) := \langle \mathcal{H}_0 \psi_0, \psi_0 \rangle = \mu_0 \langle \psi_0, \psi_0 \rangle =
\mu_0 \int_{\mathbb{R}} |\psi_0(x)|^2 dx,
\end{equation}
which is strictly positive for every eigenvalue $\mu_0$. Rewriting the spectral problem
$\mathcal{H}_0 \psi = \mu \psi$ in the Hamiltonian form
\begin{equation}
i \mathcal{H}_0 \psi = i\mu \psi,
\label{PTev-pre-Ham-last}
\end{equation}
with $\mathcal{L} = \mathcal{H}_0$, $J = i$, and $\lambda = i \mu$, we obtain the same sequence of eigenvalues
on the positive imaginary axis but associated with the positive Krein quantity (\ref{lin-krein-last}).

Of course, no contradiction is actually observed, as the Schr\"{o}dinger operator $\mathcal{H}_0$
for the quantum harmonic oscillator admits two equivalent Hamiltonian formulations
(\ref{PTev-pre-Ham}) and (\ref{PTev-pre-Ham-last}), only the former is extended
continuously with respect to the parameter $\gamma \neq 0$. In the former
formulation (\ref{PTev-pre-Ham}) with $\gamma = 0$, the self-adjoint operator $\mathcal{L} = \mathcal{P} \mathcal{H}_0$
has now two sequences of real eigenvalues: positive eigenvalues $\mu_{2N} = 4N + 1$, $N \in \mathbb{N}_0$
for the even eigenfunctions and negative eigenvalues $-\mu_{2N-1} = -4N + 1$, $N \in \mathbb{N}$ for the odd eigenfunctions.
This explains why the Krein quantity (\ref{lin-krein}) is sign-alternating even at $\gamma = 0$,
whereas the Krein quantity (\ref{lin-krein-last}) is always positive.

\section{Summary and further directions}
\label{sec-conclusion}

In the present work, we have extended the concept of the Krein signature
beyond Hamiltonian systems and applied it to $\mathcal{PT}$-symmetric systems.
We have reviewed the Hamiltonian theory, including the necessary condition
for instability bifurcation as a result of the splitting upon collision
of two eigenvalues of opposite Krein signature. An instructive case example
from the area of Bose--Einstein condensation provides a countable sequence
of nonlinear states bifurcating from eigenstates of a quantum harmonic oscillator.
The Krein signature was defined for the linearized NLS equation at each of these nonlinear states
both in the Hamiltonian and $\mathcal{PT}$-symmetric cases.
The standard properties of the Krein signature were explicitly
confirmed and the necessary condition for
instability bifurcation was verified. An illustrative (and rich in terms
of bifurcations) example was given in the form of a linear
gain/loss term in the NLS with a parabolic trap.

One can envision numerous extensions of the present theory.
On the practical side of specific applications, it would be
especially relevant to consider, e.g., two-dimensional
problems involving vorticity in settings such as the
one of~\cite{achilleos}. Also, more recently partially
$\mathcal{PT}$-symmetric settings have been introduced in~\cite{jennie,jy}
where one dimension retains the symmetry and the other dimension does
not. Considering the applicability of the ideas herein
in such systems or in systems with complex, yet
non-$\mathcal{PT}$-symmetric potentials with families
of solutions~\cite{konoextra,nixon2} would also be of interest.
Finally, from a more mathematical perspective, an understanding
of whether ideas related to the Hamiltonian-Krein theorem
can be adapted to the $\mathcal{PT}$-symmetric setting would be an
especially intriguing task.

\end{document}